\documentclass[aps,prd,twocolumn,reprint,floatfix,preprintnumbers,nofootinbib]{revtex4-2}

\usepackage{graphicx,ytableau,placeins}
\usepackage{amsmath,amssymb,amsfonts}
\usepackage[caption=false]{subfig}
\usepackage[colorlinks=True, citecolor=blue, urlcolor=blue, linkcolor=blue]{hyperref}
\newcommand{\be}{\begin{equation}}
\newcommand{\ee}{\end{equation}}
\newcommand{\ba}{\begin{eqnarray}}
\newcommand{\ea}{\end{eqnarray}}
\newcommand{\ban}{\begin{eqnarray*}}
\newcommand{\ean}{\end{eqnarray*}}
\newcommand \nn {\nonumber}
\newcommand{\rmod}{{\rm mod}~}
\newcommand{\req}[1]{({\ref{#1}})}
\usepackage{ulem}

\graphicspath{{./archive/figs/}}

\begin{document}

\title{Minimally Truncated SU(3) Lattice Gauge Theory and String Tension}
\date{\today}

\author{Vincent Chen}
\email{v.chen@duke.edu}
\affiliation{Department of Physics, Duke University, Durham, North Carolina 27708, USA}

\author{Berndt M\"uller}
\email{berndt.mueller@duke.edu}
\affiliation{Department of Physics, Duke University, Durham, North Carolina 27708, USA}

\author{Xiaojun Yao}
\email{xjyao@uw.edu}
\affiliation{InQubator for Quantum Simulation, Department of Physics,
University of Washington, Seattle, Washington 98195, USA}

\preprint{IQuS@UW-21-118}

\begin{abstract}
We study SU(3) gauge theory on small lattices in the minimal (qutrit) electric field truncation retaining only the ${\bf 1}, {\bf 3}, {\bf \overline{3}}$ representations for the link variables. Explicit expressions are given for the Kogut-Susskind Hamiltonian for the square plaquette chain and the two-dimensional honeycomb lattice. Our formalism can be easily extended to the minimally truncated general SU($N_c$) gauge theory. The addition of (static) quarks is discussed. We present results for the energy spectrum of the gauge field on these lattices by exact diagonalization of the Hamiltonian and analyze its statistical properties. We also compute the SU(3) string tension and discuss how it is modified by vacuum fluctuations. Finally, we calculate the potential energies of a static quark-antiquark pair and three static quarks and study their screening at finite temperature. 
\end{abstract}

\maketitle

\section{Introduction}
\label{sec:intro}

The goal to develop rigorous, nonperturbative methods for the computation of real-time processes in nonabelian gauge theories has motivated efforts to devise efficient quantum algorithms for Hamiltonian lattice gauge theory \cite{Byrnes:2005qx,Ciavarella:2021nmj,ARahman:2021ktn,Farrell:2022wyt,Muller:2023nnk,Illa:2024kmf,Kadam:2024ifg,Grabowska:2024emw,Balaji:2025afl,Illa:2025dou,Ciavarella:2025bsg,Jiang:2025ufg,Ciavarella:2025tdl,Li:2025sgo,Kadam:2025trs,Froland:2025bqf}. In the absence of sufficiently capable platforms for digital quantum computing for even modestly sized low-dimensional lattices, classical digital computation has been explored to compute the dynamics of SU(2) gauge fields, as a prototype of nonabelian gauge theories, on linear plaquette chains and planar hexagonal lattices with Hilbert spaces of dimensions in the range $10^4-10^5$ with the primary aim to study thermalization properties and real-time entanglement dynamics \cite{Yao:2023pht,Ebner:2023ixq,Ebner:2024mee,Ebner:2024qtu,Turro:2024pxu,Turro:2025sec,Jiang:2025ufg,Das:2025utp}. 

Here we extend these studies to the phenomenologically more interesting SU(3) gauge theory. Our work follows the electric field representation described in \cite{Byrnes:2005qx} and worked out in detail for the lowest SU(3) representations in \cite{Ciavarella:2021nmj}. As our goal is to investigate the dynamical behavior of SU(3) gauge fields on lattices comparable to those studied for SU(2), we truncate the electric field basis at the lowest nontrivial level, the qutrit truncation, which includes the singlet, triplet, and antitriplet representations. This truncation results in a Hamiltonian for a three-state Potts model, which can be generalized to SU($N_c$) gauge theory, where the Hamiltonian takes the form of a $\mathbb{Z}_{N_c}$ or ``clock'' model \cite{Ciavarella:2024fzw}. We use the shorthand notation $a \equiv_{N_c} b$ for the modular congruence $a \equiv b \, (\rmod N_c)$ throughout.

The paper is organized as follows: In Sec.~\ref{sec:SU(3)} we will discuss the SU(3) ``6j''-symbols for the simplest nontrivial truncation. The truncated SU(3) Hamiltonian on trivalent lattices including the plaquette chain and the honeycomb will be given in Sec.~\ref{sec:trivalent}, where we will also discuss the large $N_c$ limit. We will briefly explain how to insert static quarks and antiquarks at lattice sites in Sec.~\ref{sec:quarks}. Then numerical results obtained on small trivalent lattices will be shown in Sec.~\ref{sec:results}, which include spectrum, level statistics, ground state energy gap, string tension, potential energies of a static quark-antiquark pair and three static quarks and their screening at finite temperature. We will summarize and draw conclusions in Sec.~\ref{sec:conclusions}.

\section{SU(3) in the qutrit truncation}
\label{sec:SU(3)}

Irreducible representations (irreps) of SU(3) are usually denoted by their Dynkin labels $(p,q)$ representing the number of Young tableau columns with one ($p$) and two ($q$) boxes, respectively. The eigenvalues of the two Casimir operators of SU(3) are given by
\ba
C_2 &=& \frac{1}{3}(p^2+q^2+pq) + p+q \,,
\nn\\
C_3 &=& \frac{1}{18}(p-q)(3+2p+q)(3+p+2q) \,.
\ea
The simplest nontrivial truncation of the irreducible representations of SU(3) comprises the singlet ${\bf 1}=(0,0)$, the triplet ${\bf 3}=(1,0)$, and the antitriplet ${\bf\overline{3}}=(0,1)$.\footnote{For SU($N_c$) the minimal set of irreducible representations comprises all totally antisymmetric irreps, i.e.,\ the $N_c$ representations corresponding to all Young tableaux with a single column.} Using Young tableaux, these representations are depicted  as: 
\[
{\bf 1}=(0,0)\,, \qquad 
{\bf 3}=(1,0)=\ydiagram{1}\,, \qquad
{\bf\overline{3}}=(0,1)=\ydiagram{1,1} \, .
\]

Gauge invariance dictates that the electric flux lines at any lattice site combine to form a singlet \cite{Ciavarella:2021nmj}. In general, there are different ways in which three irreps (for a trivalent lattice) can be combined to a fourth. These are related by unitary transformation, which are encoded in so-called recoupling coefficients or $6j$-symbols [see \cite{Martinuzzi:2020xx} for a general analysis of the $6j$-symbols for the SU($N_c$) groups]. All valid (nonzero) $6j$-symbols must satisfy four relations among triples of the irreducible representations, called triads, which require that the singlet must be contained in the direct product of the three irreps. The four triads corresponding to the $6j$-symbol
\[
\left\{ \begin{array}{ccc} 
\lambda_1 & \lambda_2 & \lambda_3 \\  
\mu_1 & \mu_2 & \mu_3 
\end{array} \right\} \,,
\]
where $\lambda_i$ and $\mu_i$ denote SU(3) irreps $(p,q)$, are:
\be
(\lambda_1\lambda_2\lambda_3)\,,\quad
(\lambda_1\overline{\mu}_2\mu_3)\,,\quad
(\mu_1\lambda_2\overline{\mu}_3)\,,\quad
(\overline{\mu}_1\mu_2\lambda_3)\,.
\label{eq:triads}
\ee
The overline indicates the corresponding complex conjugate representation.
Corollary 1.5.2 in \cite{Martinuzzi:2020xx} states that the value of a $6j$-symbol with at least one singlet state is given by powers of $\frac{1}{\sqrt{d}}$, where $d$ describes the dimension of the specified irreducible representation. For example, since the dimensions of $\bf 3$ and $\bf\overline{3}$ are both $3$, the $6j$-symbols in the SU(3) case satisfy the following rules: 
\begin{itemize}
    \item If the valid $6j$-symbol contains only singlet states, then the $6j$-symbol equals 1. 
    \item If the valid $6j$-symbol contains three singlet states, then the $6j$-symbol equals $\frac{1}{\sqrt{3}}$.
    \item If the valid $6j$-symbol contains one or two singlet states, then the $6j$-symbol equals $\frac{1}{3}$.
    \item All other $6j$-symbols involving only the irreps $\bf 1$, $\bf 3$ and $\bf\overline{3}$ vanish.
\end{itemize}
These four rules determine all relevant $6j$-symbols for the truncated SU(3) gauge theory on trivalent lattices. Explicit values can be found in Appendix~\ref{app:6j}, as well as those for the truncated SU(4).

\section{SU(3) Hamiltonian on a trivalent lattice}
\label{sec:trivalent}

\subsection{General formalism}

\begin{figure}[t]
\centering
\includegraphics[scale=0.50]{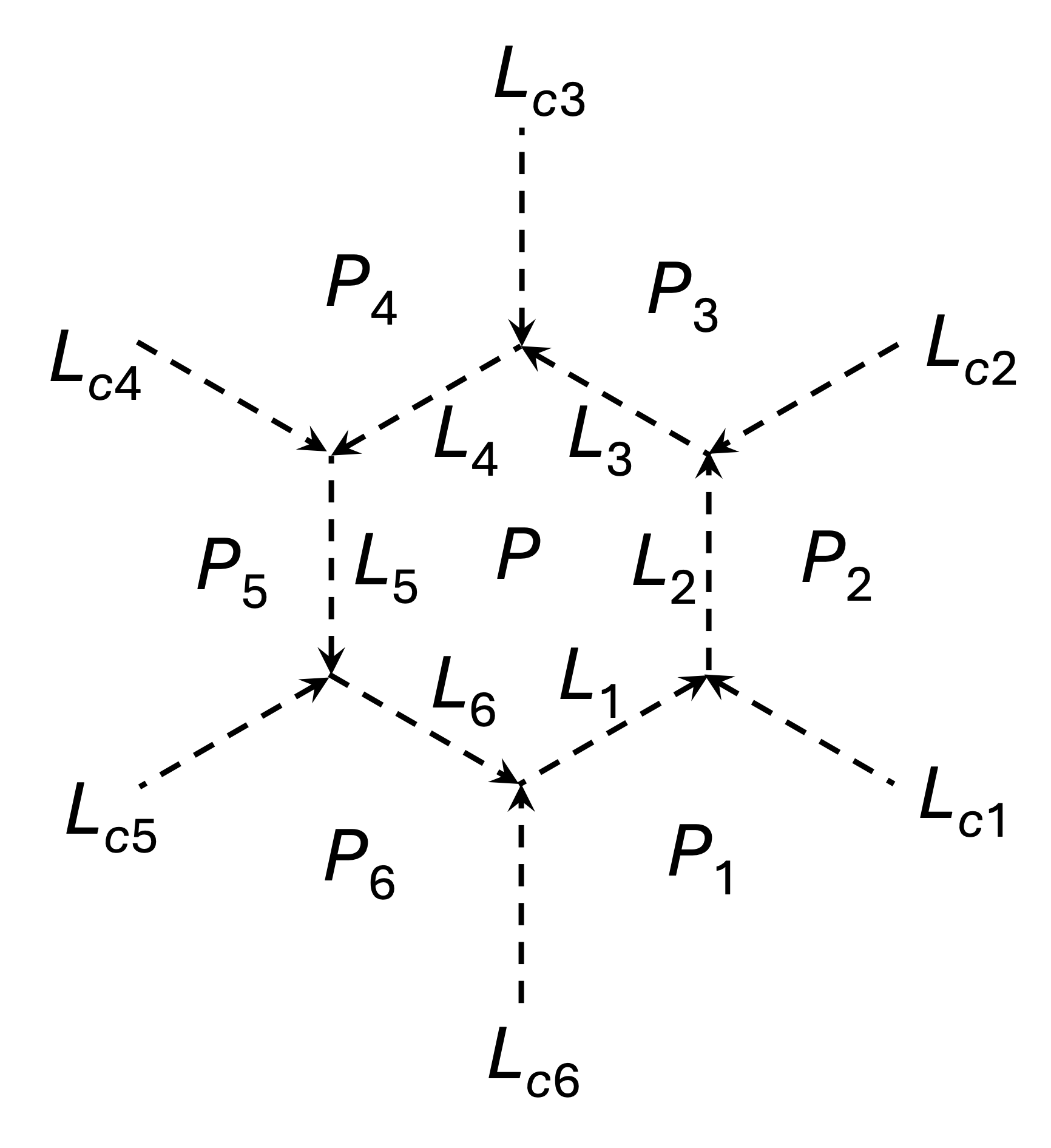}
\caption{Labeling of links and plaquettes for the hexagonal lattice. $P$ is the plaquette on which the plaquette operator acts. The arrows indicate the orientation of the links.}
\label{fig:Honeycomb}
\end{figure}

We consider the minimal truncation for SU(3) lattice gauge theory in the electric field representation given by the three lowest irreps: $({\bf 1},{\bf 3},{\bf\overline{3}})$. We label the three irreps by elements of the additive group $\mathbb{Z}_3$, corresponding to qutrits $(0,1,2)$.\footnote{The construction generalizes to all SU($N_c$) groups. In the general case it includes all irreducible configurations whose Young tableau contains only a single column, i.e.,\ all fully antisymmetric representations, which form an additive group $\mathbb{Z}_{N_c}$ under the tensor product operation.} We define the links around each plaquette in the counterclockwise (CCW) direction and label them as $L_1,\ldots,L_\nu$, where $\nu=4$ for a rectangular plaquette chain and $\nu=6$ for a planar hexagonal lattice. Our notation is illustrated for a hexagonal plaquette in Fig.~\ref{fig:Honeycomb}. We label the link endpoints as vertices $V_1,\ldots,V_\nu$. The links attaching to the vertex $V_i$ that are not sides of the plaquette under consideration are denoted as $L_{ci}$. To minimize the number of cases to be considered separately, these ``external'' links are deemed to be oriented into the vertex. i.e.,\ each vertex has two ingoing and one outgoing gauge link. For the rectangular plaquette chain, each plaquette can be thought of as being surrounded by four plaquettes, two of which (the neighboring plaquettes in the chain) are dynamical and two that remain frozen in the singlet state.

We now consider the action of the Kogut-Susskind Hamiltonian 
\be
H = \frac{g^2\Sigma}{n_l}\sum_{\rm links} (E_i^a)^2 + \frac{1}{g^2\Sigma}\sum_{\rm plaq}(6-U_P-U_P^\dagger)\,,
\label{eq:HKS}
\ee
where $n_l=2$, $\Sigma=1$ for a square plaquette, $n_l=3$, $\Sigma=3\sqrt{3}/2$ for a hexagonal plaquette, and $U_P$ denotes the corresponding plaquette operator.\footnote{Because it only causes a global shift of all energy eigenvalues by $6N_P/(g^2\Sigma)$, where $N_P$ is the number of plaquettes, we will henceforth ignore the diagonal term in the magnetic energy operator.} On a given plaquette $P$, we denote the plaquette adjacent to link $L_i$ as $P_i$, always oriented CCW consistent with the orientation of the outside link at the vertex $V_i$ (see Fig.~\ref{fig:Honeycomb}). We use the symbol $|P\rangle$ ($|P_i\rangle$) to denote the state of the plaquette in the qutrit representation: $|0\rangle = (1,0,0)$ corresponds to the electric flux ground state. Applying the plaquette operator gives $U_P |0\rangle \equiv_3 |1\rangle = (0,1,0)$, $U_P |1\rangle \equiv_3 |2\rangle = (0,0,1)$, and $U_P^3 |0\rangle = |0\rangle$ (see the Introduction for the notation $\equiv_3$; here we mean it applies to the number labeling the state). The analogous operations of the adjoint plaquette operator yield $U_P^\dagger |0\rangle = |2\rangle$, $U_P^\dagger |2\rangle = |1\rangle$, $U_P^\dagger |1\rangle = |0\rangle$. We can write all these relationships in the form:
\ba
U_P |P\rangle &\equiv_3& |P+1\rangle\, ,
\nn\\
U_P^\dagger |P\rangle &\equiv_3& |P-1\rangle\, .
\ea

We apply the same notation to the states of a link, defined with respect to the orientation of the link. If the orientation were reversed, the singlet state would remain invariant, but we would exchange the triplet and the antitriplet: $|1\rangle \leftrightarrow |2\rangle$. [This is different from the group SU(2) where the link orientation does not matter.] Using this notation, the state of a link is uniquely determined by the states of the two adjacent plaquettes:
\be
L_i \equiv_3 P - P_i \, .
\ee
Note that this relation still holds if we consider the action of the plaquette operator on plaquette $P_i$, because then $P$ and $P_i$ on the right-hand side switch their places and the orientation of the link reverses implying $L_i \to -L_i$. (Recall that $2 \equiv_3 -1$.) We can now express the states of all links connected with the vertex $V_i$ in terms of the states of the surrounding plaquettes $P,P_i,P_{i+1}$, where we identify $P_7 = P_1$:
\ba 
L_i & \equiv_3 & P - P_i \, , 
\nn\\
L_{i+1} & \equiv_3 & P - P_{i+1} \, , 
\nn\\
L_{ci} & \equiv_3 & P_i - P_{i+1} \, .
\label{eq:links}
\ea
They satisfy the relation $L_{i+1} \equiv_3 L_i + L_{ci}$\,, which is Gauss's Law.
Application of the plaquette operator to plaquette $P$ results in $L_i \to L'_i \equiv_3 L_i+1 \, $; similarly, application of the adjoint plaquette operator gives $L_i \to L'_i \equiv_3 L_i-1 \, $. The Gauss's Law $L'_{i+1} \equiv_3 L'_i + L_{ci}$ is still satisfied after the application.

We next compute the matrix elements of the plaquette operator and its adjoint on the plaquette $P$. These factorize into contributions from each of the trivalent vertices, so that it is sufficient to give the result for a single vertex $V_i$. Denoting the representations of the three links $L_i, L_{ci}, L_{i+1}$ as $r_i, r_{ci}, r_{i+1}$ with dimensions $d_i, d_{ci}, d_{i+1}$ for the initial state and with primed symbols for the final state, the matrix elements of the plaquette operator and its adjoint are obtained by repeated application of the vertex matrix elements [see \cite{ARahman:2021ktn} for a detailed derivation for SU(2)]:
\ba
M_{U_P}(V_i) &=& \sqrt{d_i d'_{i+1}} \left\{ \begin{array}{ccc} 
r_{ci} & r_i & \overline{r}_{i+1} \\  {\bf 3} & r'_{i+1} & r'_i 
\end{array} \right\} \,,
\nn\\
M_{U_P^\dagger}(V_i) &=& \sqrt{d_i d'_{i+1}} \left\{ \begin{array}{ccc} 
r_{ci} & r_i & \overline{r}_{i+1} \\  {\bf\overline{3}} & r'_{i+1} & r'_i ,
\end{array} \right\} \,,
\label{eq:vertex-ME}
\ea
where the expressions in curly brackets are the SU(3) $6j$-symbols. As explained in Sec.~\ref{sec:SU(3)}, these can only take the value $1/3, 1/\sqrt{3}$, or zero depending on the number of singlets appearing in them. Valid (nonzero) $6j$-symbols containing one or two singlets have the value $1/3$ while those with three singlets have the value $1/\sqrt{3}$. It turns out that allowed link configurations satisfying Gauss's Law involve at least one link in the singlet state, and none contain more than three singlet links. All dimension factors for the plaquette internal links are $\sqrt{3}$ except when the link is a singlet.

As noted in Sec.~\ref{sec:SU(3)}, for the $6j$-symbol $\left\{ \begin{array}{ccc} \lambda_1 & \lambda_2 & \lambda_3 \\  \mu_1 & \mu_2 & \mu_3 \end{array} \right\}$ to be non-vanishing it has to satisfy the four triad conditions \eqref{eq:triads}. In order to see how our vertex rules satisfy these constraints, we employ the $\mathbb{Z}_3$ notation for the link states $L_i$ instead of the standard notation for their SU(3) representations $r_i$. We then can express the matrix elements \req{eq:vertex-ME} as
\be 
M_\Pi(V_i) = \sqrt{d_i d'_{i+1}} \left\{ \begin{array}{ccc} 
L_{ci} & L_i & -L_{i+1} \\  \Pi & L'_{i+1} & L'_i 
\end{array} \right\} \,,
\label{eq:Vertex}
\ee
where $\Pi = 1$ for the plaquette operator and $\Pi = -1 \equiv_3 2 \, $ for its adjoint. The triad rules then take the form:
\begin{align}
L_{ci}+L_i-L_{i+1} &\equiv_3 0\,,&  
L_{ci}-L'_{i+1}+L'_i &\equiv_3 0\,,
\nn\\
\Pi+L_i-L'_i &\equiv_3 0\,, & 
-\Pi+L'_{i+1}-L_{i+1} &\equiv_3 0\,.
\end{align}
These constraints coincide with the relations noted below \req{eq:links}: $L'_i \equiv_3 L_i+\Pi \, $ and the Gauss's Law constraint before and after the action of the plaquette operator: $L_{i+1} \equiv_3 L_i+L_{ci} \, $ and $L'_{i+1} \equiv_3 L'_i+L_{ci} \, $.

Since the initial and final states of the links are uniquely fixed by the states of the active plaquette and its six neighbors, all vertex matrix elements can be expressed in terms of the plaquette states. These can easily be derived from the relations \req{eq:links} and \req{eq:vertex-ME}. For each combination of external plaquettes $P_i, P_{i+1}$ the matrix elements of the combined operator $(U_P + U_P^\dagger)$ in the magnetic part of the lattice Hamiltonian can be written as a $3\times 3$ matrix $M_{P,P'}^{(P_i,P_{i+1})}$ for the initial and final active plaquette states with vanishing diagonal elements. 
Table \ref{tab:Vertex-ME} lists these $3\times 3$ matrices for a single trivalent vertex. The matrices are asymmetric because of the asymmetry of the definition \eqref{eq:Vertex}.
\begin{table}[tb]
\begin{tabular}{c|ccc}
& 0 & 1 & 2 \\
\hline
 0\, & $\left(
 \begin{array}{ccc}
 0 & 1 & 1 \\
 1 & 0 & 1 \\
 1 & 1 & 0 \\
\end{array}
\right)$ & $\left(
\begin{array}{ccc}
 0 & 1 & 1 \\
 \frac{1}{3} & 0 & \frac{1}{\sqrt{3}} \\
 \frac{1}{\sqrt{3}} & 1 & 0 \\
\end{array}
\right)$ & $\left(
\begin{array}{ccc}
 0 & 1 & 1 \\
 \frac{1}{\sqrt{3}} & 0 & 1 \\
 \frac{1}{3} & \frac{1}{\sqrt{3}} & 0 \\
\end{array}
\right)$ \\
 & & & \\
 1\, & $\left(
\begin{array}{ccc}
 0 & \frac{1}{3} & \frac{1}{\sqrt{3}} \\
 1 & 0 & 1 \\
 1 & \frac{1}{\sqrt{3}} & 0 \\
\end{array}
\right)$ & $\left(
\begin{array}{ccc}
 0 & 1 & 1 \\
 1 & 0 & 1 \\
 1 & 1 & 0 \\
\end{array}
\right)$ & $\left(
\begin{array}{ccc}
 0 & \frac{1}{\sqrt{3}} & 1 \\
 1 & 0 & 1 \\
 \frac{1}{\sqrt{3}} & \frac{1}{3} & 0 \\
\end{array}
\right)$ \\
 & & & \\
 2\, & $\left(
\begin{array}{ccc}
 0 & \frac{1}{\sqrt{3}} & \frac{1}{3} \\
 1 & 0 & \frac{1}{\sqrt{3}} \\
 1 & 1 & 0 \\
\end{array}
\right)$ & $\left(
\begin{array}{ccc}
 0 & 1 & \frac{1}{\sqrt{3}} \\
 \frac{1}{\sqrt{3}} & 0 & \frac{1}{3} \\
 1 & 1 & 0 \\
\end{array}
\right)$ & $\left(
\begin{array}{ccc}
 0 & 1 & 1 \\
 1 & 0 & 1 \\
 1 & 1 & 0 \\
\end{array}
\right)$ \\
\end{tabular}
\caption{The matrix elements $M_{P,P'}^{(P_i,P_{i+1})}$ \eqref{eq:Vertex} of the sum of the plaquette operator and its adjoint at a single trivalent vertex $V_i(L_i,L_{i+1};L_{ci})$ for different combinations of the neighboring plaquette states. It is assumed that the links around the active plaquette are oriented counterclockwise, and that the external links are oriented toward the vertex, as shown in Fig.~\ref{fig:Honeycomb}. The rows are labeled by the quantum state $P_i$ of the external plaquette adjacent to the incoming link $L_i$; the columns indicate the state of the plaquette $P_{i+1}$ adjacent to the outgoing link $L_{i+1}$. The rows and columns of the $3\times 3$ matrices denote the initial and final quantum states $(P,P')$ of the active plaquette in the order $(0,1,2)$.}
\label{tab:Vertex-ME}
\end{table}

Similarly, the full plaquette matrix elements of the magnetic energy term in the Hamiltonian can be arranged into $3\times 3$ matrices solely on the basis of the qutrit plaquette states. The number of independent $3\times 3$ matrices is greatly reduced by the rotational and reflection symmetries of the plaquette geometry as we will discuss in the next two subsections.

\subsection{Linear plaquette chain}

\begin{figure}[h]
\centering
\includegraphics[width=0.9\linewidth]{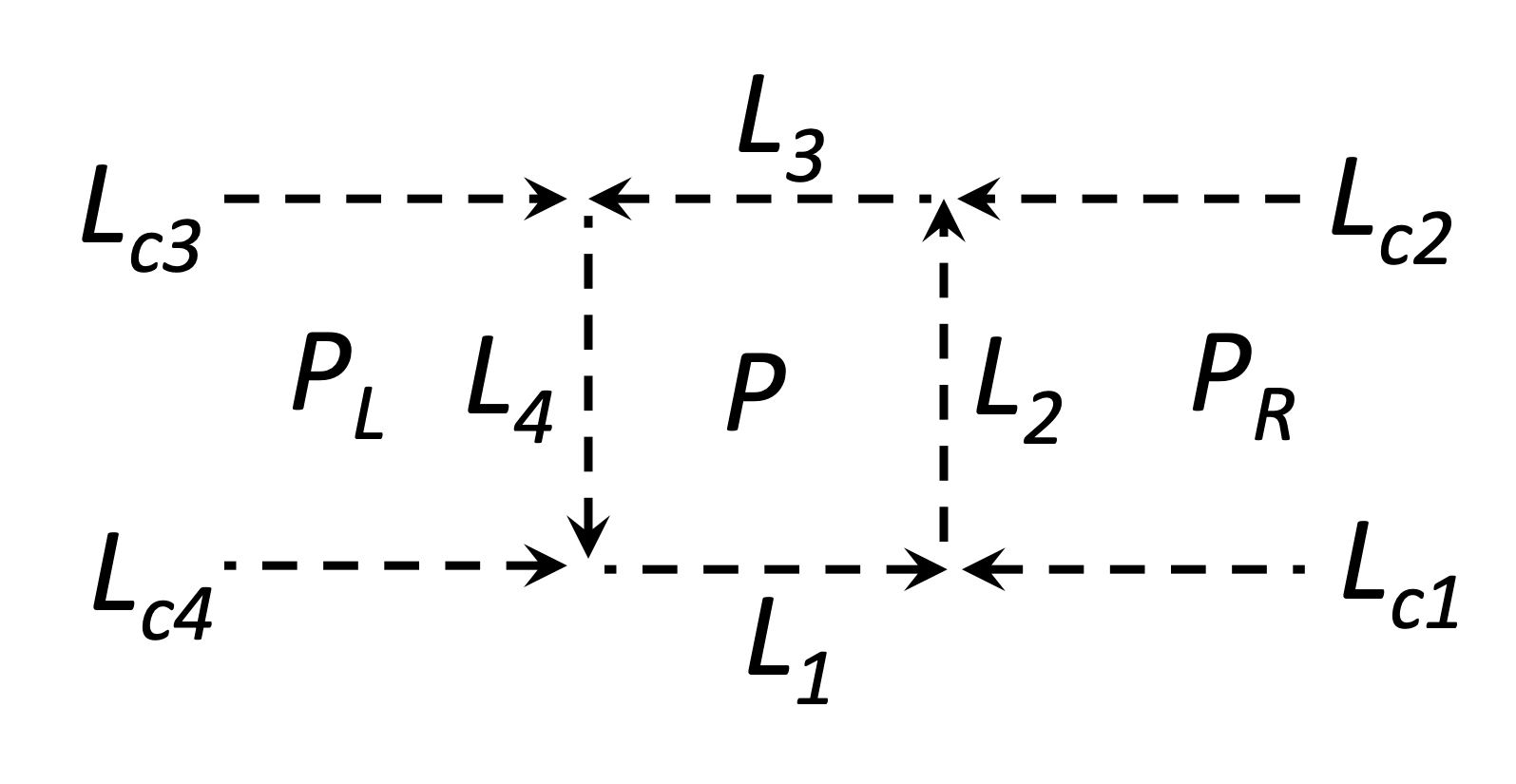}
\caption{Structure of an active plaquette $P$ in the linear plaquette chain.}
\label{fig:plaq-chain}
\end{figure}

Previous work has considered the simplest truncation for the linear plaquette chain~\cite{Ciavarella:2021nmj} by imposing Gauss's law on qutrit link states. Here we consider plaquette excitations as a qutrit per plaquette, which also works for the full two-dimensional space on the honeycomb lattice. On the linear chain the active plaquette $P$ has two adjacent plaquettes, one to the right and one to the left, which we denote by $P_R$ and $P_L$, respectively, as shown in Fig.~\ref{fig:plaq-chain}. We label the four vertices CCW starting from the lower right vertex: $V_1,\ldots,V_4$, and the four links beginning with the bottom link, as $L_1,\ldots,L_4$, again CCW. 

In order to apply the general formalism we can define adjacent plaquette states for all four links as $P_1=P_B, P_2=P_R, P_3=P_T, P_4=P_L$ (bottom, right, top, left) with $P_B=P_T=|0\rangle$.\footnote{Different choices for $P_B$ and $P_T$ describe the presence of global electric flux through the plaquette chain. We will discuss this aspect further in Sec.~\ref{sec:string} on the SU(3) string constant.} The states of the four links of the active plaquette can then be expressed in terms of the plaquette states as 
\begin{align}
L_1 &=\ L_3 = P \,, 
\nn\\ 
L_2 & \equiv_3  P - P_R \, , 
\nn\\ 
L_4 & \equiv_3  P - P_L \, .
\end{align}
and the states of the four external links are given by
\begin{align}
-L_{c1} & \equiv_3  L_{c2} = P_R  \, , 
\nn\\ 
-L_{c3} & \equiv_3  L_{c4} = P_L  \, .
\end{align}
Here again the convention is to orient the external links into the vertices. One easily checks that the Gauss's Law relations $L_i+L_{ci} = L_{i+1}$ are trivially fulfilled. The action of the magnetic term in the Hamiltonian can now be obtained by repeatedly applying the formula \req{eq:Vertex}. Explicit values of the matrix elements are given in Table \ref{tab:Hmag-chain} in Appendix~\ref{app:MPP'-chain}. The resulting Hamiltonian can also be written as a spin-1 quantum Potts model, as shown in Appendix~\ref{app:6j}.

\subsection{Planar hexagonal lattice}

The matrix elements of a hexagonal plaquette shown in Fig.~\ref{fig:Honeycomb} are easily calculated using the general expression \eqref{eq:Vertex} consecutively for each of the six vertices, but are too many independent different configurations of adjacent plaquette states for the hexagonal lattice to be listed explicitly. To find the number of independent configurations, we note that for the qutrit truncation each plaquette within the planar lattice can be completely specified by the ordered sequence of the states of its edge links, numbered 1 through 6 counterclockwise. Let this collection of numbers, alternatively ${\bf 1}$, ${\bf 3}$, or $\bf\overline{3}$ or $0,1,2$, be called $X$. To calculate the number of unique sequences among these $3^6 = 729$ configurations, accounting for the symmetries of the hexagon, we can use Burnside's lemma. Since the symmetries of the hexagon are represented by the group dihedral group $D_6$ the number of distinct configurations is:
\be
|X/D_6| = \frac{1}{|D_6|} \sum_{g\in D_6}|X^g|,
\ee
where $|D_6| = 12$ denotes the order of $D_6$ and $X^g = \{ x\in X\, |\, g\cdot x = x\}$ is the set of elements in $X$ fixed by a group element $g$.
Then, the following list describes the number of elements of $X$ fixed by each element of $D_6$: 
\begin{enumerate}
    \item All $3^6$ elements are fixed by the identity element.
    \item $3$ elements are fixed by the two 60 degree rotations.
    \item $3^2$ elements are fixed by the two 120 degree rotations. 
    \item $3^3$ elements are fixed by the 180 degree rotation.
    \item $3^3$ elements are fixed by the three reflections about a line bisecting two opposite sides of the hexagon.
    \item  $3^4$ elements are fixed by the three reflections about a line connecting opposite vertices of the hexagon.
\end{enumerate}
Therefore, the number of distinct sequences is given by:
\be
|X/D_6|=\frac{1}{12}(3^6 + 2 \cdot 3 + 2 \cdot 3^2 +  3^3 + 3\cdot 3^3 + 3 \cdot 3^4) = 92 \,.
\ee

\subsection{Large $N_c$ limit}
\label{sec:largeNc}

It is interesting to consider higher SU($N_c$) gauge groups and the limit of a large number of colors $N_c$, which has been investigated before and applied to the SU(3) gauge theory \cite{Ciavarella:2024fzw}. In the case of SU(4), the minimal truncation including all antisymmetric irreducible representations comprises the irreps ${\bf 1}$, ${\bf 4}$, ${\bf 6}$, and $\bf\overline{4}$. The relevant $6j$-symbols are given in Table \ref{tab:6jsu4} in Appendix~\ref{app:6j}; they can be obtained by application of the rules for base SU($N_c$) $6j$-symbols given by Martinuzzi \cite{Martinuzzi:2020xx} and in \cite{Alcock-Zeilinger:2022hrk}. 

We now discuss which matrix elements of the plaquette operator survive in the large-$N_c$ limit. Following the arguments outlined in \cite{Ciavarella:2024fzw} this is best achieved by considering the electric flux lines on the active plaquette and its neighbors. The plaquette operator $U_P$ adds a CCW flux line around the active plaquette; its adjoint $U_P^\dagger$ adds a CW flux line. Each CCW flux line has quantum numbers in the fundamental representation; each CW flux line has quantum numbers in the antifundamental representation. 

The matrix elements of the plaquette operator are determined by the relative weight of the resulting representations of the links shared with the neighboring plaquettes. Consider, e.g., a shared link in the $k$-tensor antisymmetric irrep $|k\rangle$, corresponding to a Young tableau with $k$ boxes in a single column [Dynkin label $(0^{k-1},1,0^{N_c-k-1})$], which might be generated by the action of $(U_P^\dagger)^k$ on the adjacent plaquette. Acting with $U_P$ adds one unit of electric flux, either antisymmetrically resulting in the irrep $|k+1\rangle$, or symmetrically resulting in the mixed symmetry irrep represented by the Young diagram with $k$ boxes in the first column and one box in the second column [Dynkin label $(1,0^{k-2},1,0^{N_c-k-1})$]. The dimension of the antisymmetric combination is $D_{\rm a}=\tbinom{N_c}{k+1}$; that for the symmetric combination is $D_{\rm s} = k\tbinom{N_c+1}{k+1}$. One easily checks that $D_{\rm a} + D_{\rm s} = N_c\tbinom{N_c}{k}$, giving the antisymmetric combination the relative weight 
\be
w_{\rm a}(k,+) = \frac{D_{\rm a}}{D_{\rm a} + D_{\rm s}} = \frac{N_c-k}{(k+1)N_c} ,
\ee
which approaches $1/(k+1)$ in the large-$N_c$ limit. The matrix element of $U_P$ then acquires a factor $\sqrt{w_{\rm a}(k,+)}$. An analogous consideration applies to the shared link on the other side of the plaquette. For example, for $N_c=3$ and $k=1$, one obtains $w_{\rm a}(1,+) = 1/3$, contributing a factor $1/\sqrt{3}$ to the matrix element.

On the other hand, the application of the operator $U_P^\dagger$ reduces the flux on the shared link by one unit, yielding $D_{\rm a} = \tbinom{N_c}{k-1}$ and $D_{\rm s} = k(N_c^2-N_c(k-1)-k)\tbinom{N_c}{k-1}$. In this case the relative weight of the antisymmetric combination is
\be
w_{\rm a}(k,-) = \frac{k}{(N_c-k+1)N_c} ,
\ee
which is suppressed by two powers of $N_c$ in the large-$N_c$ limit. 

This implies that in the large-$N_c$ limit only those gauge field configurations survive, in which a flux carrying plaquette is surrounded by plaquettes with zero or counter-circulating electric flux. Note that our result differs somewhat from that obtained in \cite{Ciavarella:2021nmj}. Of course, as plaquettes with increasing electric flux are considered, the minimal truncation becomes an increasingly unreliable approximation to SU($N_c$) as more and more configurations with higher flux representations on links are neglected.

\section{Quarks}
\label{sec:quarks}

\begin{figure}[htb]
\centering
\includegraphics[width=0.6\linewidth]{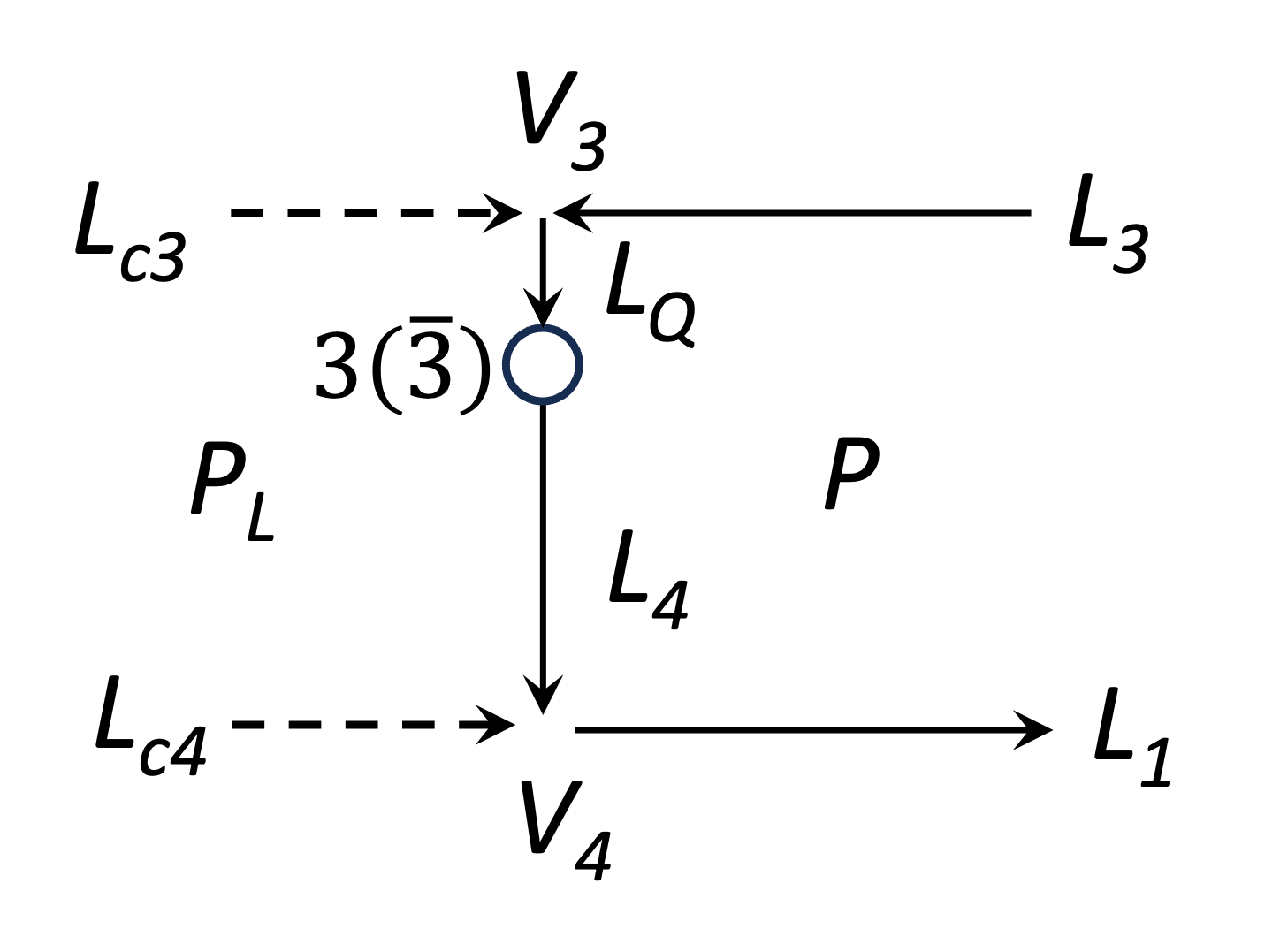}
\caption{Insertion of a static quark or antiquark at vertex $V_3$ by shifting the insertion point infinitesimally away from the lattice site and creating an auxiliary link $L_Q$ of infinitesimal length and a new trivalent vertex $V_Q(L_Q,L_4;1)$ (quark) or $V_Q(L_Q,L_4;-1)$ (antiquark). The links of the active plaquette $P$ are shown as solid arrows, and those of the adjacent plaquette $P_L$ as dashed arrows.}
\label{fig:split}
\end{figure}

Quarks and antiquarks are represented by triplet and antitriplet operator insertions at the lattice sites, respectively. In order to avoid having to deal with the complications arising from the need to combine four irreps into a singlet to implement gauge invariance, one can use ``point splitting'' to maintain the trivalent vertex structure. The most convenient way to do so is to offset the (static) quark insertion point infinitesimally away from the lattice site to a point on the connected vertical link, as indicated in Fig.~\ref{fig:split}.
Altogether, for the example shown in the figure, the insertion replaces the vertex $V_3(L_3,L_4;L_{c3})$ with the product $V_3(L_3,L_Q;L_{c3})V_Q(L_Q,L_4;1)$ for $\bf 3$ and $V_3(L_3,L_Q;L_{c3})V_Q(L_Q,L_4;-1)$ for $\bf\overline{3}$. Because the auxiliary link is of infinitesimal length, it does not contribute to the electric energy or to the plaquette area. For the minimal truncation considered here, the group representation of the auxiliary link $L_Q$ is uniquely determined by $L_4$ and the SU(3) quantum numbers of the insertion; for higher truncations more possibilities arise. Furthermore, since the vertical link is shared between two plaquettes, the analogous modification occurs when $P_L$ is the active plaquette.

As the (anti)quark insertion changes the color flow within the lattice by injecting electric flux in the (anti)fundamental representation, gauge invariance dictates that the injected flux is terminated at another insertion: The flux injected by a quark must terminate on an antiquark, and vice versa. When the source and sink are located $n$ lattice spacings $a$ apart, this implies the existence of an additional electric flux over a distance $na$. When the lattice chain is in its ground state at strong bare coupling with most links in the singlet representation, this must raise the energy of the lattice by an amount proportional to the quark-antiquark separation, $\Delta E_{Q\bar{Q}} \approx \frac{2}{3}g^2n$, where we set $a=1$. When the lattice is in a highly excited state before the $Q\bar{Q}$ pair insertion, the energy may not increase because the addition of electric flux is a vector addition in group space and may not increase the average flux on the links connecting the quark and antiquark. These considerations suggest that the $Q\bar{Q}$ potential is confining in the lattice ground state, but may be screened in a highly excited state of the lattice. We further discuss the SU(3) string tension for the plaquette chain and honeycomb in Sec.~\ref{sec:string} where we present an alternative way to compute it that does not require the introduction of explicit quark sources.

\section{Results}
\label{sec:results}

\subsection{Spectrum}
\subsubsection{Linear plaquette chain}
\label{sec:chain}

\begin{figure}[h]
\centering
\includegraphics[width=0.9\linewidth]{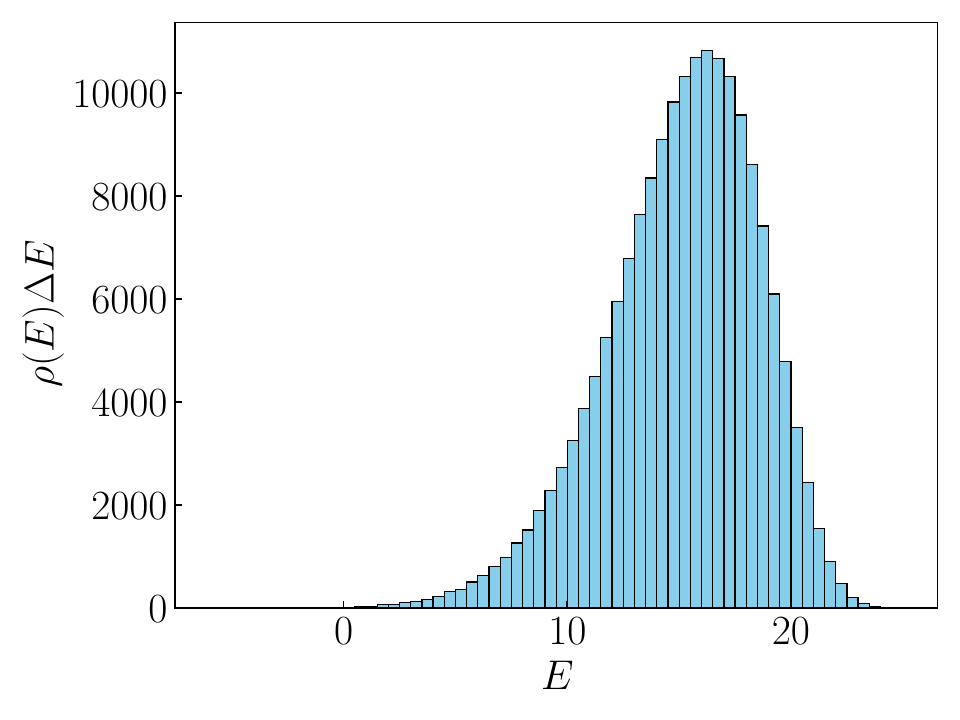}
\caption{Level spectrum $\rho(E)\Delta E$ for the $N=11$ plaquette chain with asymmetric (singlet/triplet) boundary conditions at $g^2=1$ with the bin size $\Delta E=0.5$. The energies are shown without the diagonal term in the magnetic energy. Compared with the spectrum for an SU(2) chain (see Fig. 1 of \cite{Ebner:2023ixq}) the level density for the SU(3) chain is more strongly skewed to the upside.}
\label{fig:spectrum}
\end{figure}

We first discuss some results for the linear plaquette chain, beginning with the level statistics for $N=11$ plaquettes with fixed boundary conditions. In order to avoid level degeneracies, we choose asymmetric boundary conditions with external plaquettes in the $|0\rangle$ and $|1\rangle$ states, respectively, at the left and right ends of the chain. The level spectrum for $g^2=1$ is shown in Fig.~\ref{fig:spectrum}.

\begin{figure}[t]
\centering
\includegraphics[width=0.9\linewidth]{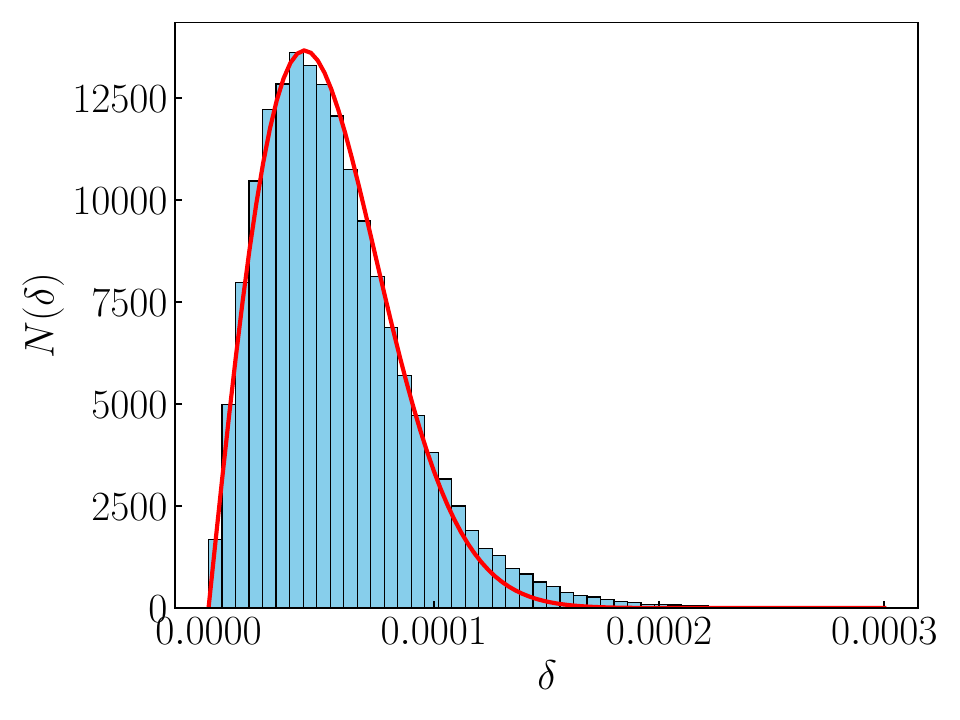}
\caption{Energy gap distribution $N(\delta)$ for the middle $70\%$ spectrum (see Fig.~\ref{fig:spectrum}) of the $N=11$ plaquette chain with asymmetric (singlet/triplet) boundary conditions at $g^2=1$. The bin size is $6\times10^{-6}$. The red curve is a fit of the Wigner surmise form $N(\delta) = a\delta\,e^{-b\delta^2}$ with $a=5.2894\times10^8$ and $b=2.7554\times10^8$. The distribution is well described by a Wigner-Dyson distribution for the GOE ensemble.}
\label{fig:gap-dist}
\end{figure}

We then study the level statistics. The energy gap is defined as $\delta_\alpha = E_\alpha - E_{\alpha-1}$ with $E_\alpha$ denoting the eigenenergy ordered from small to large. If a quantum system is ergodic, its level spacings are expected to follow the Wigner-Dyson distribution for the Gaussian Orthogonal Ensemble (GOE), featuring level repulsion. We show the distribution of level spacings for the $N=11$ plaquette chain at $g^2=1$ in Fig.~\ref{fig:gap-dist}, where the lowest and highest $15\%$ eigenenergies are removed since states near the spectrum edges are more sensitive to finite volume effects. The distribution is well described by the Wigner-Dyson statistics. 

\begin{figure}[t]
\centering
\includegraphics[width=0.9\linewidth]{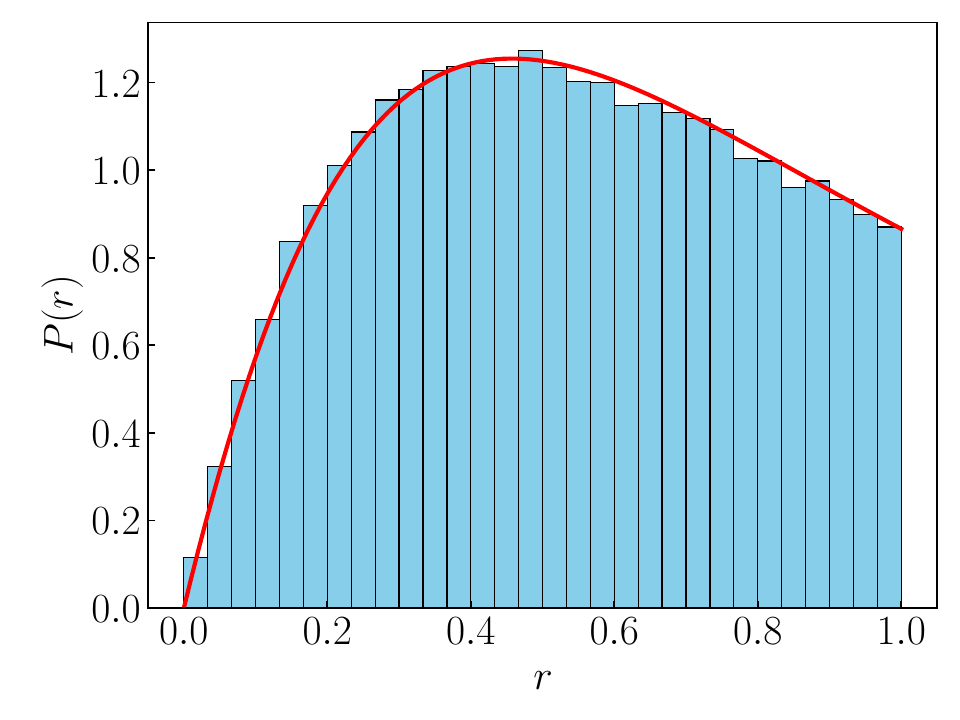}
\caption{Normalized energy gap ratio distribution for the middle $70\%$ spectrum of the $N=11$ plaquette chain with asymmetric (singlet/triplet) boundary conditions at $g^2=1$. The bin size is $1/30$. Expected gap ratio distribution \eqref{eq:GOEgap} for a GOE ensemble is shown as the red solid line.}
\label{fig:gap-ratio}
\end{figure}

Figure~\ref{fig:gap-ratio} shows a histogram of the normalized energy gap ratio 
\be
r_\alpha = \frac{{\rm min}[\delta_\alpha,\delta_{\alpha-1}]}{{\rm max}[\delta_\alpha,\delta_{\alpha-1}]} \le 1 \,,
\ee
 together with the analytical result \cite{Jansen:2019}
\be
P_{\rm GOE}(r) = \frac{27}{4}\frac{r+r^2}{(1+r+r^2)^{5/2}} \,,
\label{eq:GOEgap}
\ee
for GOE level spectrum (solid red line). The average gap ratio $\langle r \rangle = 0.5317$ is in good agreement with the GOE prediction $\langle r \rangle_{\rm GOE} = 0.5307$ \cite{Atas:2013gvn}, confirming the quantum chaotic nature of the SU(3) plaquette chain for this coupling.

\begin{figure}[t]
\centering
\includegraphics[width=0.9\linewidth]{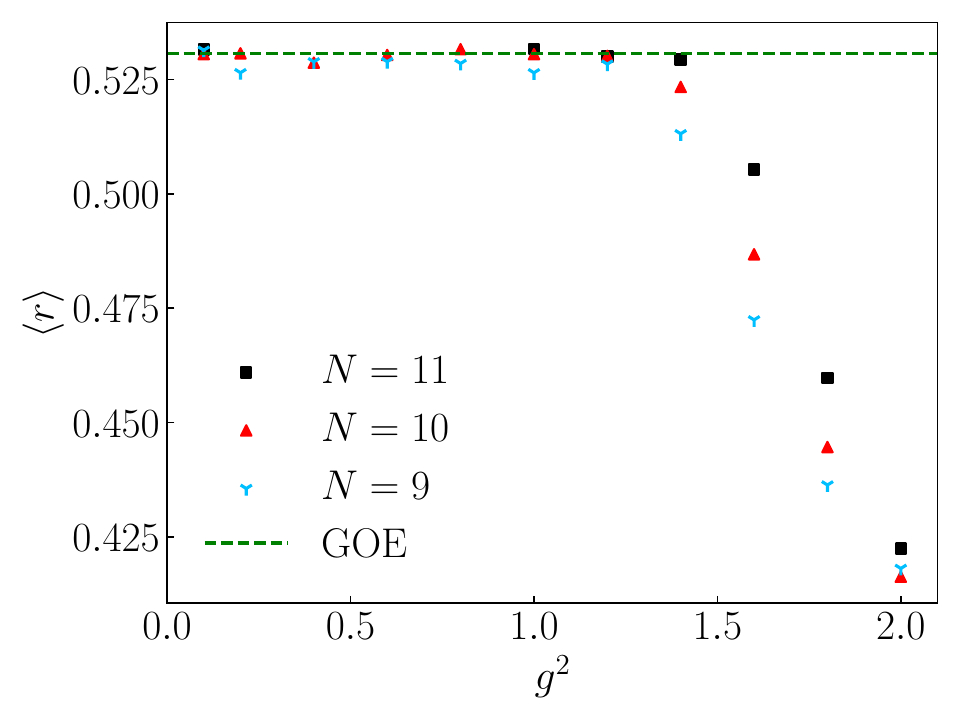}
\caption{Average value of gap ratios as a function of coupling on plaquette chains of different lengths. As the coupling increases, a longer chain is needed to fully exhibit quantum chaos.}
\label{fig:gap-ratio-g2}
\end{figure}

Using the average value of gap ratios as an indicator of quantum chaos, we identify the coupling range for chaotic SU(3) plaquette chains in Fig.~\ref{fig:gap-ratio-g2}. At couplings $g^2\leq 1.2$, a $N=9$ plaquette chain is already approximately chaotic. Further increasing the chain length reduces statistical uncertainties and makes the agreement with the GOE prediction even better. At larger couplings, a short plaquette chain is not chaotic. Enlarging the system size increases the nonintegrability and the level of quantum chaos. The larger the coupling, the longer the chain needs to be to fully exhibit quantum chaos.

\begin{figure}[h]
\centering
\includegraphics[width=0.9\linewidth]{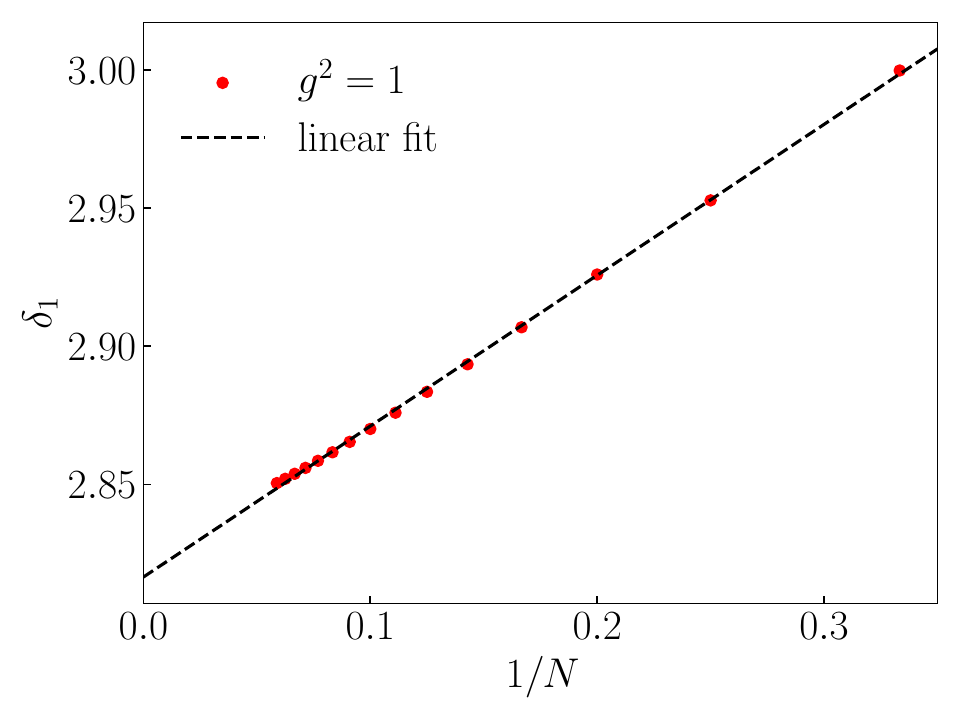}
\caption{Ground state energy gaps (red points) for plaquette chains with $N\in[3,11]$ for confining boundary conditions at $g^2=1$. The dashed black line is a linear fit against the inverse plaquette length.}
\label{fig:GSgap}
\end{figure}

Finally, we study the dependence of the ground state energy gap for the linear plaquette chain with symmetric confining boundary conditions (external singlet plaquettes on both ends of the chain) as a function of chain length $N \in [3,17]$. As Fig.~\ref{fig:GSgap} shows, the ground state gap scales linearly with the inverse length of the plaquette chain, extrapolating to the value $\lim_{N\to\infty}\delta_1 = 2.8165$.

\subsubsection{Planar hexagonal lattice}
\label{sec:hexagon}

\begin{figure}[h]
\subfloat[Parallelogram.\label{fig:honeycomb_parallelogram}]{%
  \includegraphics[height=1.2in]{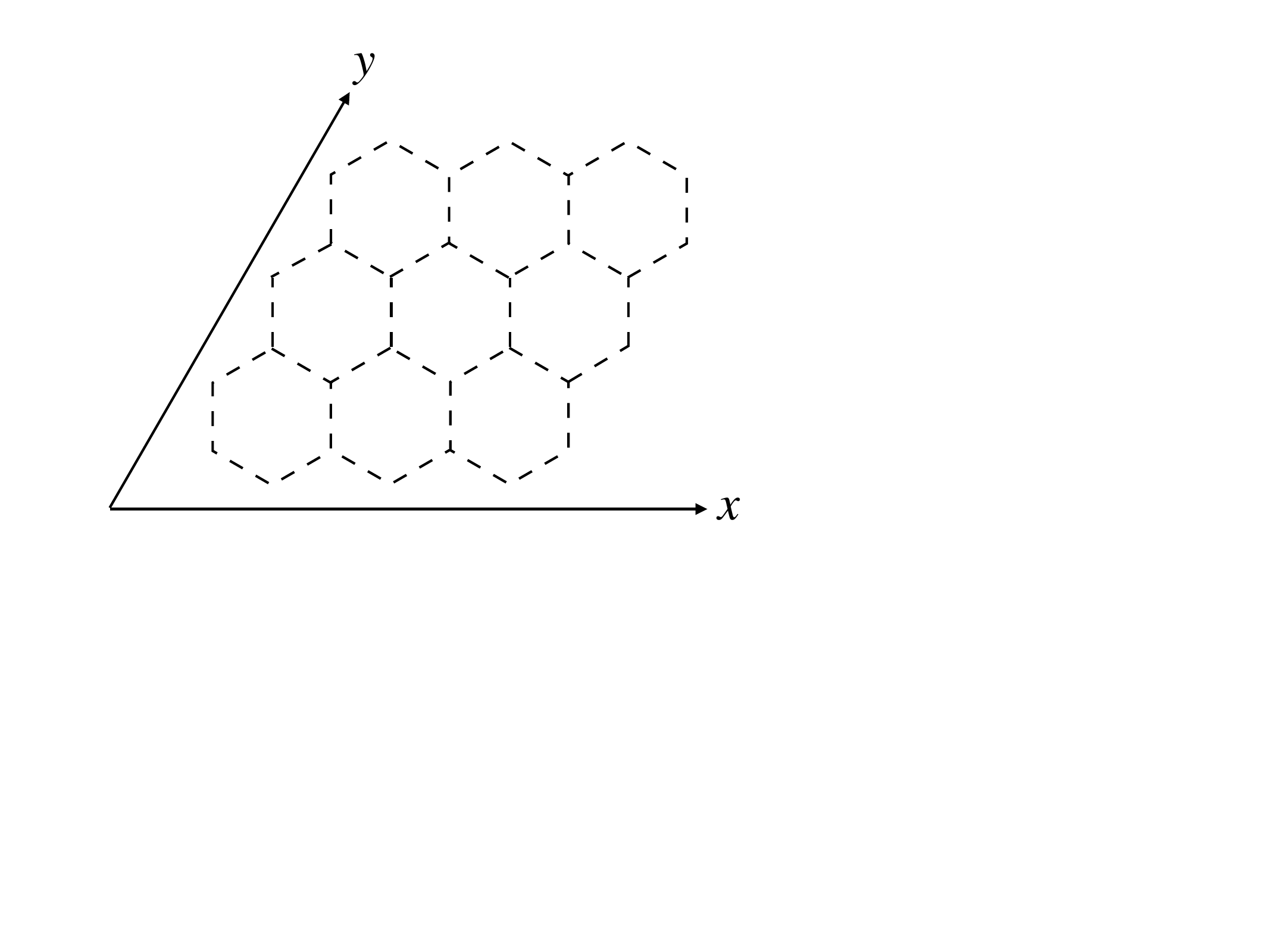}%
}\hfill
\subfloat[Approximate rectangle.\label{fig:honeycomb_rectangle}]{%
  \includegraphics[height=1.2in]{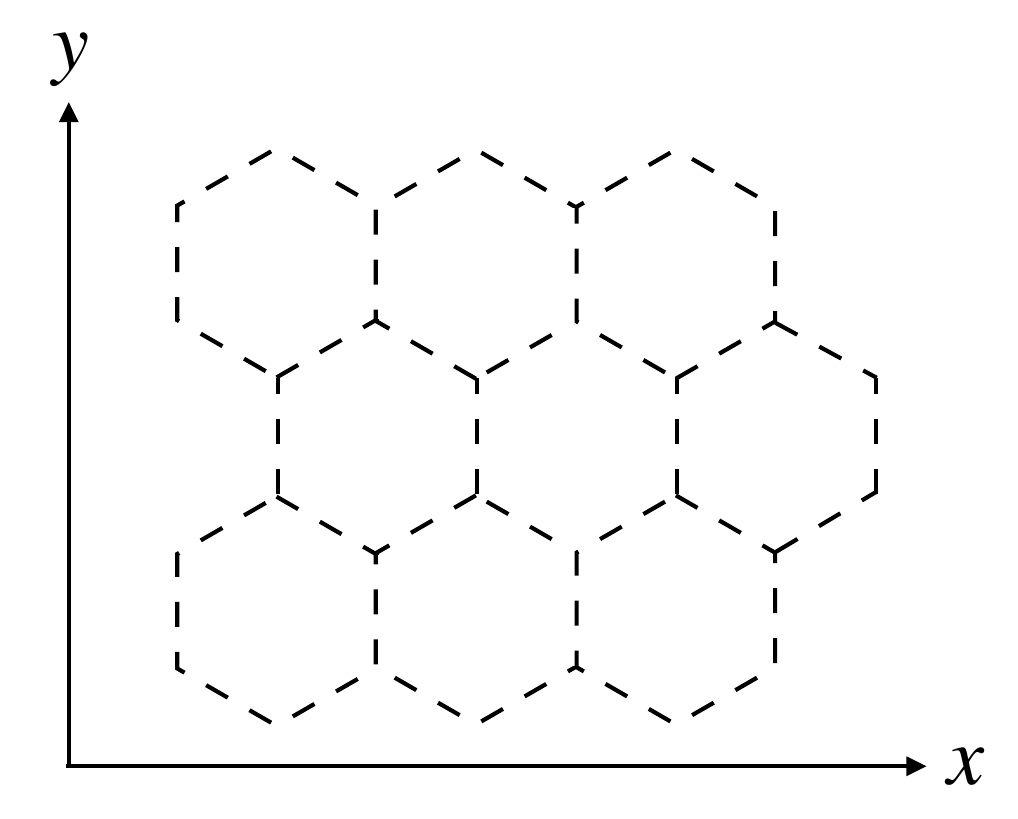}%
}
\caption{Examples of the honeycomb lattice considered in this work: (a) a parallelogram with $N_x=N_y=3$; (b) an approximate rectangle with $N_x=N_y=3$.}
\label{fig:honeycomb_lattice}
\end{figure}

A finite planar hexagonal (``honeycomb'') lattice can have different shapes. In this work, we consider two shapes: parallelogram and approximate rectangle shown in Fig.~\ref{fig:honeycomb_lattice}. The lattice size is characterized by two numbers $N_{x},N_{y}$ indicating the number of plaquettes in the two planar directions oriented at $60^\circ$ and  $90^\circ$ with respect to each other for the parallelogram and approximate rectangle, respectively. 

For the studies of spectrum and level statistics, we use the parallelogram shape. The boundary conditions can either be periodic or fixed with external plaquette states $P_T,P_B,P_R,P_L$ at the top, bottom, left, and right edges of the parallelogram spanned by the $N_P = N_x\times N_{y}$ active plaquettes. To avoid degeneracies, we here choose $P_T=1, P_B=0, P_L=0, P_R=2$.\footnote{If there is an ambiguity in whether an external plaquette is at the top/bottom, or left/right, we assign it to be at the top/bottom.} 

\begin{figure}[t]
\centering
\includegraphics[width=0.9\linewidth]{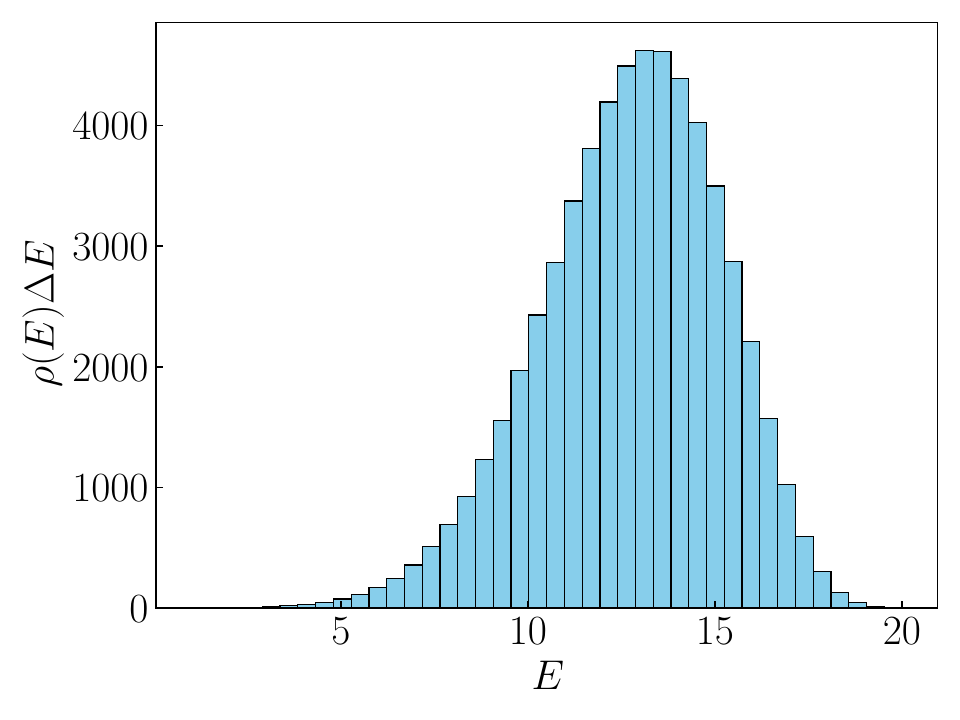}
\caption{Level spectrum $\rho(E)\Delta E$ for the $5\times 2$ planar hexagonal lattice with mixed singlet/triplet/antitriplet boundary conditions at $g^2\Sigma=1$. The bin size is $\Delta E=0.5$. Compared with the spectrum for the plaquette chain (see Fig.~\ref{fig:spectrum}) the level density for the SU(3) honeycomb lattice theory is less strongly skewed on the upside.}
\label{fig:spectrum-hex}
\end{figure}

A histogram of the energy spectrum of the $5\times 2$ hexagonal lattice is shown in Fig.~\ref{fig:spectrum-hex} at the gauge coupling $g^2=1/\Sigma \approx 0.3849$, where $\Sigma$ is the hexagon area defined below \eqref{eq:HKS}. One notices that the spectrum is slightly less skewed to the upside than that for the plaquette chain. 

\begin{figure}[h]
\centering
\includegraphics[width=0.9\linewidth]{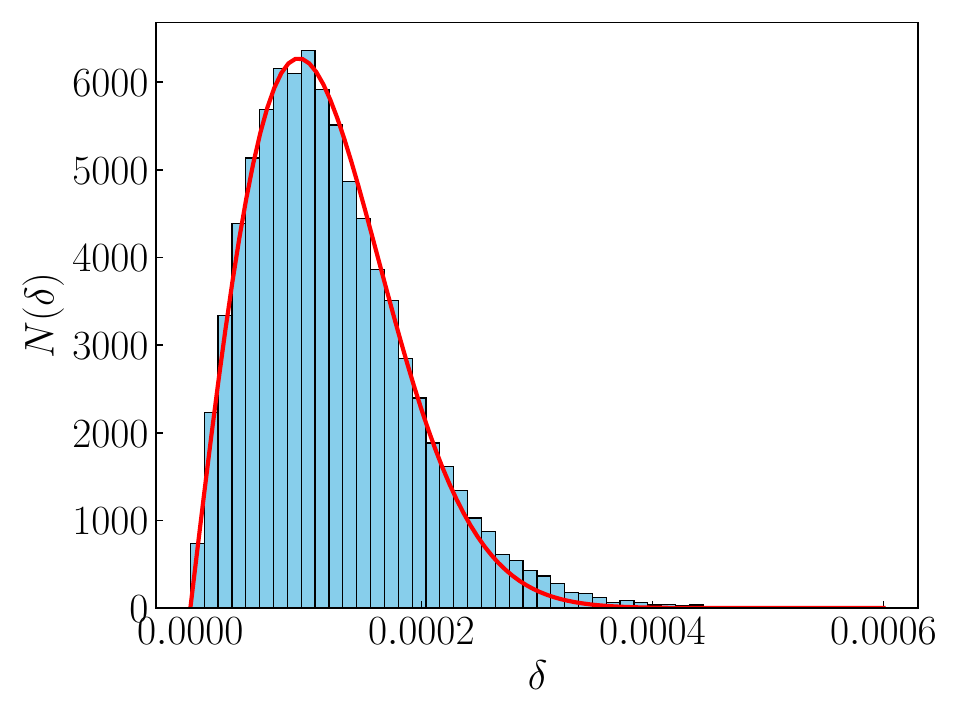}
\caption{Energy gap distribution $N(\delta)$ for the middle $70\%$ spectrum (see Fig.~\ref{fig:spectrum-hex}) of the $5\times 2$ hexagonal lattice with mixed singlet/triplet/antitriplet boundary conditions at $g^2\Sigma=1$. The bin size is $1.2\times10^{-5}$. The red curve is a fit of the Wigner surmise form $ax\,e^{-bx^2}$ with $a=1.1025\times10^8$ and $b=5.6857\times10^7$. The distribution is well described by a Dyson-Wigner distribution for the GOE ensemble.}
\label{fig:gap-dist-hex}
\end{figure}

Figure \ref{fig:gap-dist-hex} shows the energy gap distribution for the middle $70\%$ spectrum of the $5\times 2$ hexagonal lattice with mixed boundary conditions shown in Fig.~\ref{fig:spectrum-hex} together with a Wigner-Dyson fit. The distribution clearly exhibits the strong nearest-neighbor level repulsion characteristic for quantum chaotic systems described by the GOE ensemble.

\begin{figure}[h]
\centering
\includegraphics[width=0.9\linewidth]{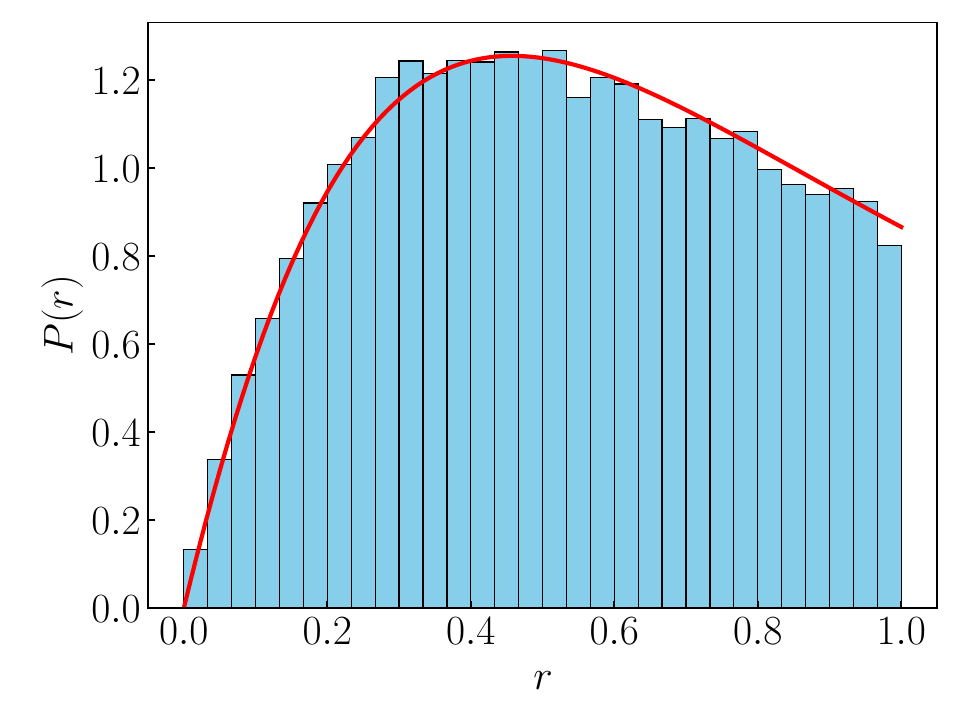}
\caption{Normalized energy gap ratio distribution for the $5\times 2$ hexagonal lattice with mixed singlet/triplet/antitriplet boundary conditions at $g^2\Sigma=1$. Red solid line shows the expected gap ratio distribution \eqref{eq:GOEgap} for a GOE ensemble.}
\label{fig:gap-ratio-hex}
\end{figure}

Figure~\ref{fig:gap-ratio-hex} shows a histogram of the normalized energy gap ratio $P(r)$ together with the analytical result $P_{\rm GOE}(r)$ for the GOE level spectrum (solid red line). The average gap ratio $\langle r \rangle = 0.5303$ is in good agreement with the GOE prediction $\langle r \rangle_{\rm GOE} = 0.5307$, confirming the quantum chaotic nature of the SU(3) hexagonal lattice for this coupling.

\begin{figure}[t]
\centering
\includegraphics[width=0.9\linewidth]{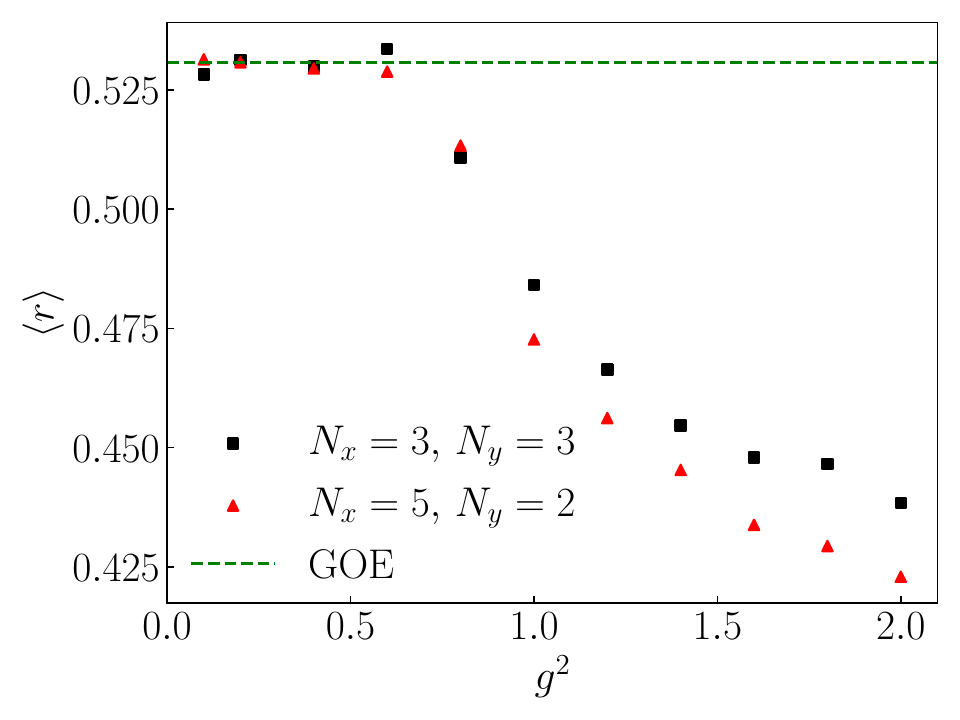}
\caption{Average value of gap ratios as a function of coupling on honeycomb lattices of different sizes. At small couplings, small systems already exhibits quantum chaos approximately. At large couplings, the seven-body interaction term on the smaller $3\times3$ lattice is more effective in integrability breaking than the fewer-body terms on the bigger $5\times2$ lattice.}
\label{fig:gap-ratio-g2-hex}
\end{figure}

To study the coupling dependence of the quantum chaotic nature on finite lattices, we plot the average value of gap ratios as a function of coupling in Fig.~\ref{fig:gap-ratio-g2-hex}. As in the case of the plaquette chain, small systems already exhibit quantum chaotic behavior at weak couplings. What differs from the plaquette chain is the coupling value where the system starts to be more integrable. If we take $g^2\Sigma$ as the effective coupling that corresponds to the coupling $g_s^2$ on square lattice (including the case of plaquette chain), we find the coupling value at which the system starts to deviate from quantum ergodicity is about the same, $1.3$. Furthermore, we note that at large couplings, the bigger $5\times2$ lattice is actually more integrable than the smaller $3\times3$ lattice. We attribute to the fact that the magnetic term in the Hamiltonian breaks integrability and the central plaquette operator on the $3\times3$ lattice is a seven-body interaction term, which is absent on the $5\times2$ lattice. We expect the seven-body term to be more effective in the integrability breaking than the fewer-body terms on the $5\times2$ lattice.

\begin{figure}[t]
\centering
\includegraphics[width=0.9\linewidth]{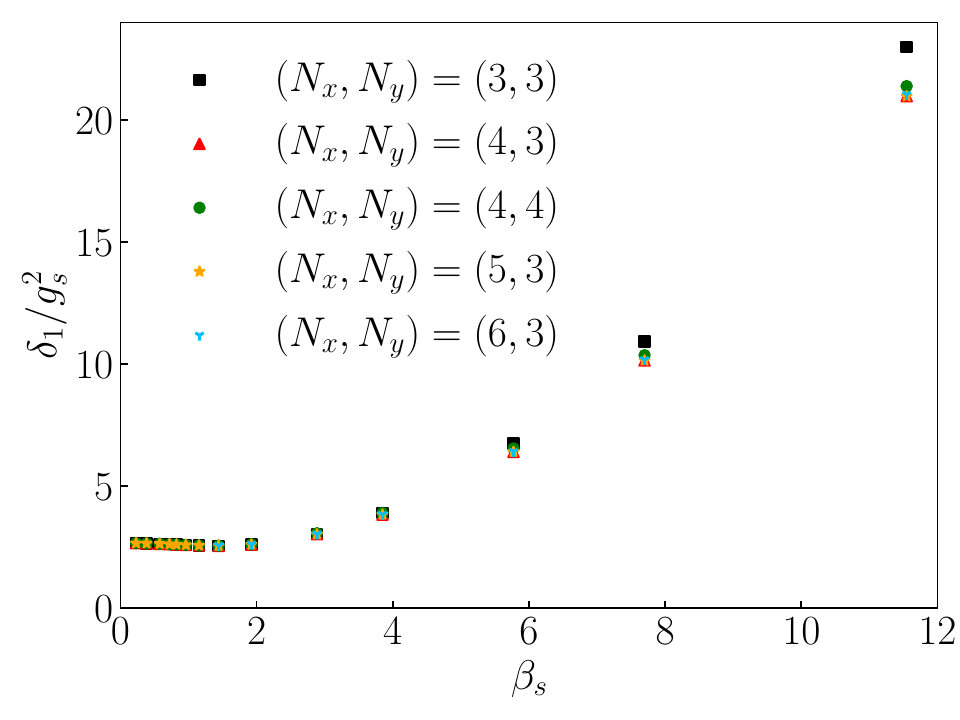}
\caption{Ground state energy gap normalized by the effective coupling squared $g_s^2$ as a function of $\beta_s=N_c/g_s^2$ for hexagonal lattices of different sizes.}
\label{fig:gap-g2-hex}
\end{figure}

\begin{figure}[h]
\centering
\includegraphics[width=0.9\linewidth]{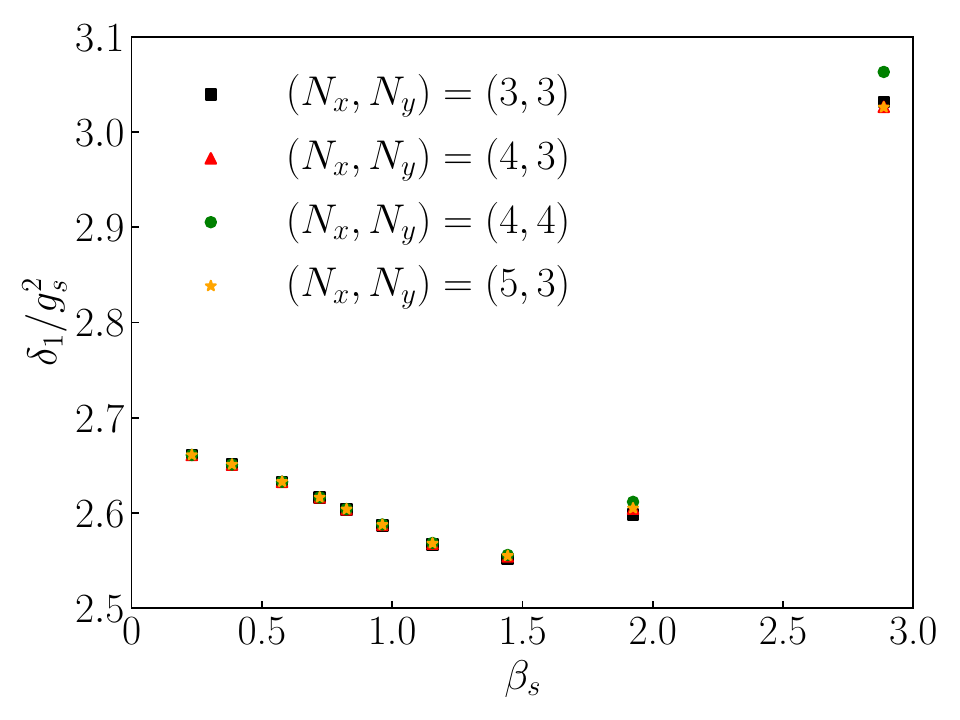}
\caption{Zoom-in of Fig.~\ref{fig:gap-g2-hex} in the small $\beta_s$ regime. The asymptotic value at strong coupling agrees with Fig.~6 of~\cite{Carlsson:2003wx}. The valley shape at small $\beta_s$ agrees qualitatively too.}
\label{fig:gap-g2-hex-small-beta}
\end{figure}

Finally, we study the ground state gap by using the approximate rectangle shape.
The relation between the ground state gap normalized by the coupling squared and the coupling is shown in Fig.~\ref{fig:gap-g2-hex}, with a zoom-in of the strong coupling regime in Fig.~\ref{fig:gap-g2-hex-small-beta}. In order to compare with previous Euclidean studies of $(2+1)$D SU(3) lattice gauge theory, we introduce $g_s^2 = g^2\Sigma$, which is the effective coupling on the hexagonal lattice corresponding to the one on the square lattice, as explained above, and $\beta_s = N_c/g^2_s = 3/g_s^2$. At strong coupling, i.e., small $\beta_s$, the asymptotic value agrees with that in Fig.~6 of~\cite{Carlsson:2003wx}, and the trend as $\beta_s$ increases agrees qualitatively, though the valley location and and depth differ quantitatively. We attribute this to the truncation effect, as higher irreps are more important at weaker couplings. At small couplings, the magnetic term dominates in the truncated SU(3) theory and we expect $\delta_1 \propto 1/g_s^2$, which is confirmed by the quadratic increase in Fig.~\ref{fig:gap-g2-hex}. In general, it is known that in the large $\beta_s$ regime, finite size effects are significant for the ground state gap but can be cured by using a linear lattice size of order $10$ (see Fig.~6 of~\cite{Carlsson:2003wx}).

\subsection{SU(3) string tension and melting}
\label{sec:string}
\subsubsection{Linear plaquette chain}
\label{sec:string-chain}

A simple method of measuring the SU(3) string tension is to inject fundamental electric flux into the plaquette chain at one end and extract it at the other end. This can be achieved by choosing the plaquette state $P_T$ adjacent to link $L_3$ (i.e., on the top of $P_L$ and $P_R$) in Fig.~\ref{fig:plaq-chain} in the antitriplet state, or $P_T = |2\rangle$ in the qutrit notation. The boundary condition injects triplet electric flux via the external link $L_{c2}$ at the upper right corner of the chain. If we set the external plaquettes $P_L=P_R=|0\rangle$, the flux is extracted at the upper left corner. Alternatively, if we choose $P_L=|2\rangle$, the flux is extracted via the external link at the lower left corner. 

We can think of this field configuration as due to a static quark and antiquark attached to opposing ends of the lattice. Because the gauge field on the plaquette chain is fully dynamical, the chromoelectric flux can take any path through the lattice; the Hamiltonian \eqref{eq:HKS} couples all possible configurations of electric flux consistent with the boundary conditions. We now show that the energy difference between the two alternative boundary conditions vanishes exponentially with the length of the plaquette chain.

\begin{figure}[h]
\centering
\includegraphics[width=0.9\linewidth]{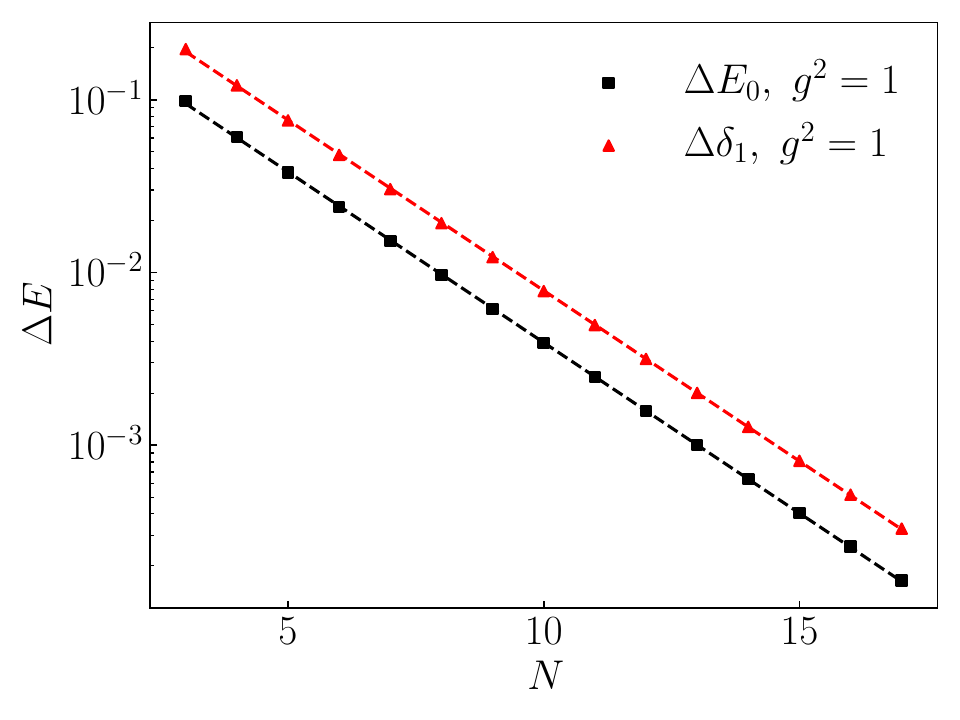}
\caption{Chain length dependence of the differences in the ground state energy (black squares) and its gap (red triangles) between the two flux penetration configurations where the triplet flux is (i) injected and extracted at the two upper corners of the plaquette chain and (ii) injected and extracted at two diagonally opposed corners. The ground state energy and gap differences are seen to fall exponentially with the chain length $N$ as $\Delta E_0 \approx 0.3723\,e^{-0.4551N}$ and $\Delta \delta_1 \approx 0.7431\,e^{-0.4550N}$, respectively for $g^2=1$, shown by the fitted dashed lines.  The identical fall-off confirms that the fluctuation mode that flips the top and bottom horizontal fluxes becomes soft as the length of the chain increases.
}
\label{fig:E-split}
\end{figure}

The minimal path for the electric flux to flow from the upper right corner to the lower left corner of the chain is longer by one lattice spacing, and thus one expects the energy for the diagonally flowing flux to be larger than that for the flux that flows straight through from one upper corner to the other one. Quantum fluctuations in the ground state smear this picture out, because in both cases flux loops around individual plaquettes cause the injected flux to wander around randomly between the upper and lower edges of the plaquette chain. As a result, the energy difference between the two boundary configurations decreases exponentially with the length of the chain as shown by the black square points in Fig.~\ref{fig:E-split}. The same exponential fall-off is seen in the difference of the ground state energy gap in the presence of the electric flux for the two different boundary conditions (red triangle points) in Fig.~\ref{fig:E-split}. 

\begin{figure}[h]
\centering
\includegraphics[width=0.9\linewidth]{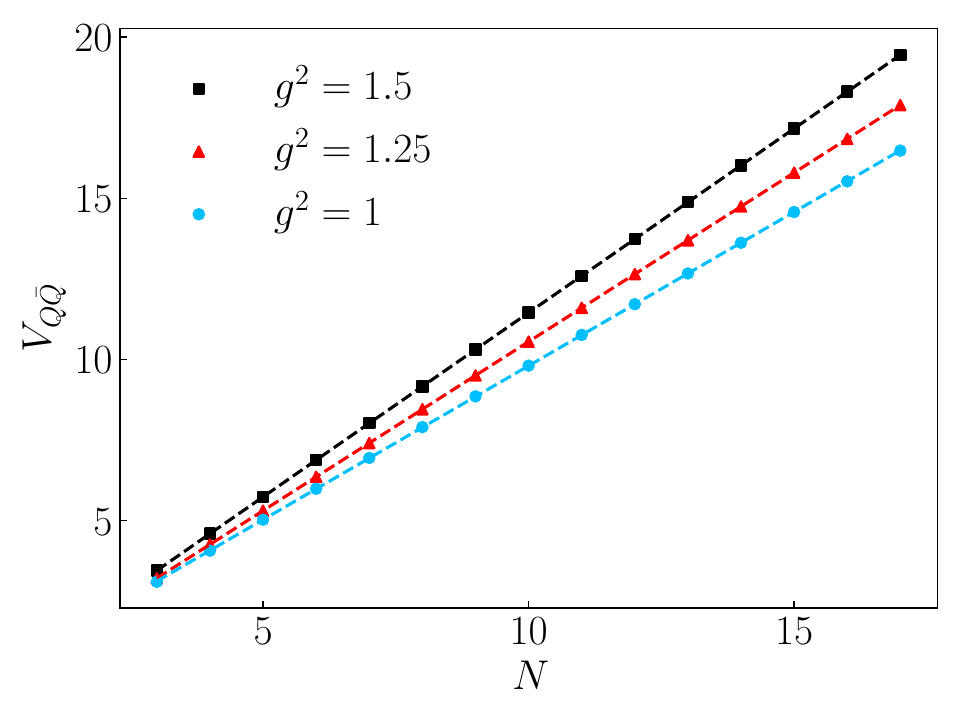}
\caption{Chromoelectric flux tube energy $V_{Q\bar{Q}}$ on the linear plaquette chain as a function of length $N$ for the coupling constants $g^2=1.5$ (black squares), $g^2=1.25$ (red triangles) and $g^2=1$ (blue discs). The linear slope is the fundamental string tension $\sigma$ of SU(3) on a linear plaquette chain.}
\label{fig:VQQ}
\end{figure}

Figure \ref{fig:VQQ} shows the difference in the ground state energy between the flux-penetrated (upper right corner to upper left corner) and flux-free lattice. We associate this difference with the potential energy $V_{Q\bar{Q}}$ of a heavy quark pair attached to the two ends of the plaquette chain. As the lattice theory has a mass gap and is confined, we expect $V_{Q\bar{Q}}$ to grow linearly with the chain length $N$. This is borne out by explicit computation as shown in Fig.~\ref{fig:VQQ} for three different couplings: $g^2=1.5$ (black squares), $g^2=1.25$ (red triangles) and $g^2=1$ (blue discs). 

\begin{figure}[h]
\centering
\includegraphics[width=0.9\linewidth]{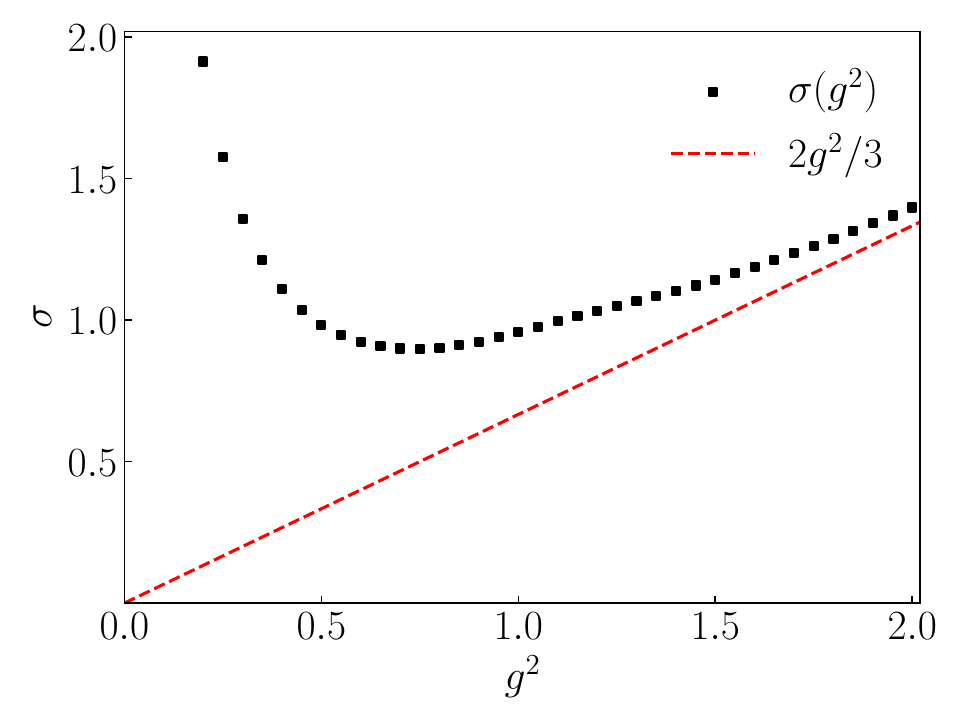}
\caption{Coupling constant dependence of the string constant $\sigma$ of SU(3) on a linear plaquette chain.}
\label{fig:sigma}
\end{figure}

The dependence of the fundamental string tension $\sigma$ on the coupling constant is depicted in Fig.~\ref{fig:sigma}. The strong coupling limit result $\sigma_0=\frac{2}{3}g^2$ is shown as a dashed red line for comparison, which is just the electric energy of an excited link. The deviation from the strong coupling limit value reflects the effect of vacuum fluctuations of the gauge field. The magnetic term in the gauge field Hamiltonian \eqref{eq:HKS} excites local electric flux loops which modify the shortest (straight) path of the flux through the lattice. As the coupling weakens, more and more local flux loops are excited, and the flux path begins to meander through the chain. The deviation between the black points and the dashed red line can be interpreted as the average increase in the length of path along which the electric flux flows. This deviation can be understood as a renormalization of the bare SU(3) string tension $\sigma_0$. The minimum string tension $\sigma_{\rm min}\approx 0.9$ occurs around $g^2 \approx 0.75$.

\begin{figure}[h]
\centering
\includegraphics[width=0.9\linewidth]{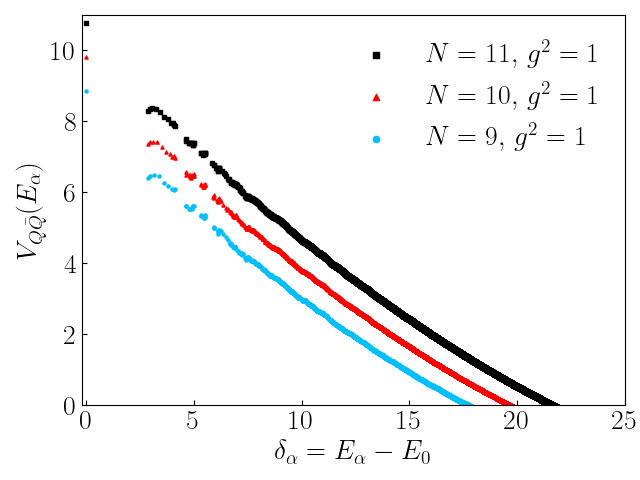}
\caption{Dependence of the $Q\bar{Q}$ potential $V_{Q\bar{Q}}(E_\alpha)$ on the excitation energy of the energy eigenstate $E_\alpha$ of the flux-free lattice. The curves are for $N=9\,,10\,,11$ and coupling $g^2=1$.}
\label{fig:VQQvsE}
\end{figure}

\begin{figure}[h]
\centering
\includegraphics[width=0.9\linewidth]{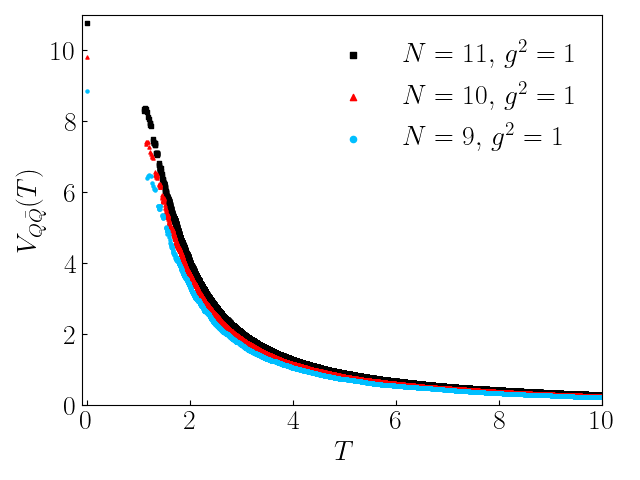}
\caption{Canonical temperature dependence of the $Q\bar{Q}$ potential $V_{Q\bar{Q}}(T)$ for the $N=9\,,10\,,11$ plaquette chains at coupling $g^2=1$.}
\label{fig:VQQvsT}
\end{figure}

Also, we explore the dependence of $V_{Q\bar{Q}}$ on the excitation energy of the state. In Fig.~\ref{fig:VQQvsE} we show the energy difference between eigenstates of the flux-free lattice ($E_\alpha$) and those of the lattice in the presence of a global fundamental chromoelectric flux along the top [$E_\alpha^{(Q\bar{Q})}$]
\be
V_{Q\bar{Q}}(E_\alpha) = E_\alpha^{(Q\bar{Q})} - E_\alpha \,,
\ee
as a function of the excitation energy $E_\alpha-E_0$ of the state for the flux-free $N = 9\,,10\,,11$ plaquette chains at $g^2=1$. Clearly, the potential energy associated with the global flux decreases with increasing energy and approaches zero when the energy eigenvalue reaches the peak of the spectral density $\rho(E)$. The spectral peak corresponds to infinite microcanonical temperature $T(E) = (d\ln\rho(E)/dE)^{-1}$ or infinite canonical temperature $E = \sum_\alpha E_\alpha e^{-E_\alpha/T}/\sum_\alpha e^{-E_\alpha/T}$ in the thermodynamic limit,  indicating that the confining $Q\bar{Q}$ potential ``melts'' at high temperature. The resulting potential energy as a function of the canonical temperature $V_{Q\bar{Q}}(T)$ is shown in Fig.~\ref{fig:VQQvsT}. As the figure shows, there is no threshold temperature above which the string tension vanishes. While the pure two-dimensional SU(3) gauge theory exhibits a deconfinement transition in the termodynamic limit \cite{Christensen:1990qs,Engels:1996dz,Liddle:2008kk} as does the planar three-state Potts model \cite{Wu:1982ra}, it is unknown whether the linear plaquette chain in the qutrit truncation shows a similar phase transition.

\subsubsection{Planar hexagonal lattice}
\label{sec:string-hexagon}

\begin{figure}[t]
\subfloat[$Q\bar{Q}$.\label{fig:QQ}]{%
  \includegraphics[height=1.2in]{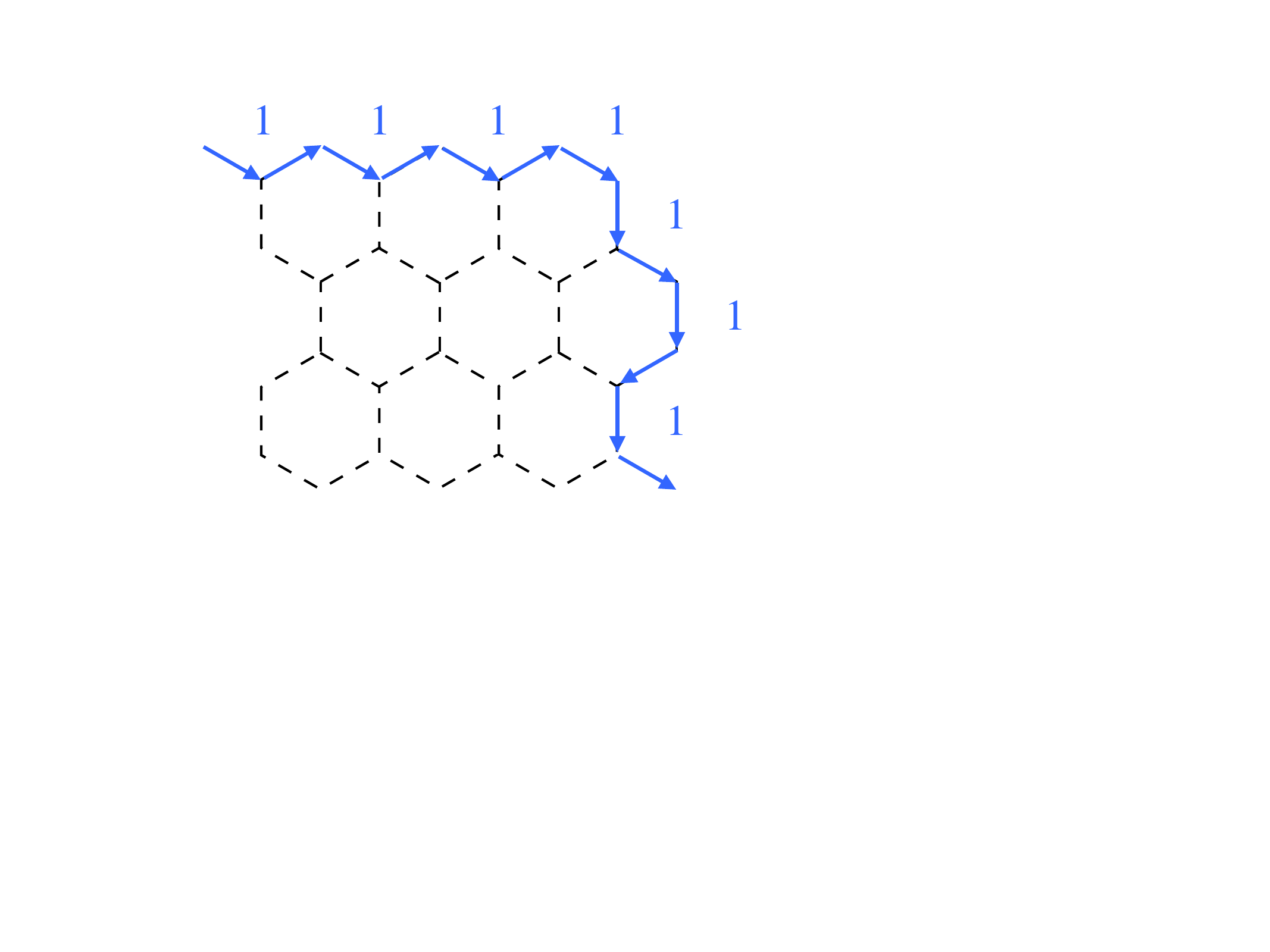}%
}\hfill
\subfloat[$QQQ$.\label{fig:QQQ}]{%
  \includegraphics[height=1.2in]{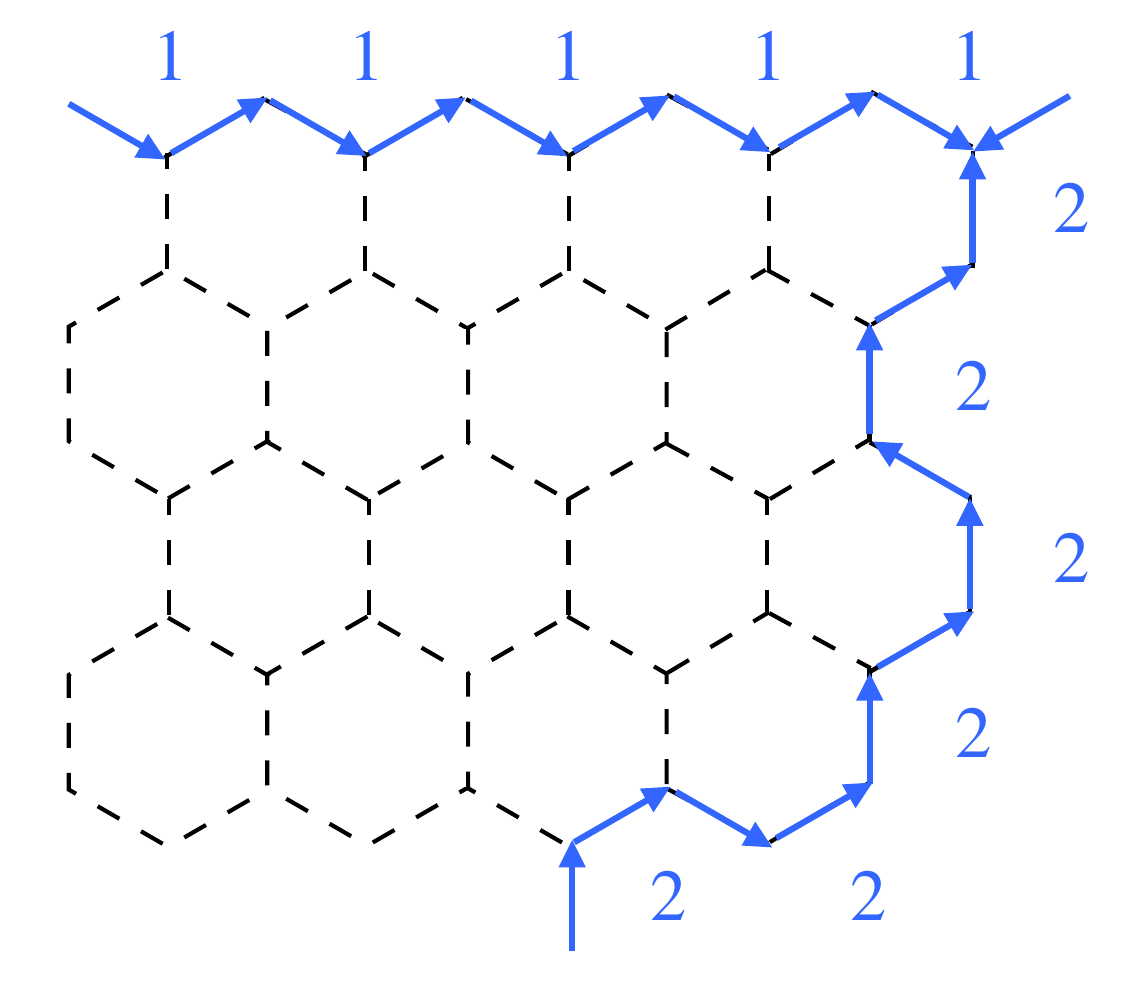}%
}
\caption{Hexagonal lattices in approximate rectangle shape with injected electric fluxes. (a) Injected fundamental and antifundamental fluxes at the diagonal ends. (b) Three injected fundamental fluxes. The numbers indicate the states of external plaquettes as the boundary conditions.}
\label{fig:external_Q}
\end{figure}

Next we study the string tension on a hexagonal lattice. We use the approximate rectangle shape and introduce incoming and outgoing electric fluxes along the diagonal direction, as shown in Fig.~\ref{fig:QQ}.

\begin{figure}[h]
\centering
\includegraphics[width=0.9\linewidth]{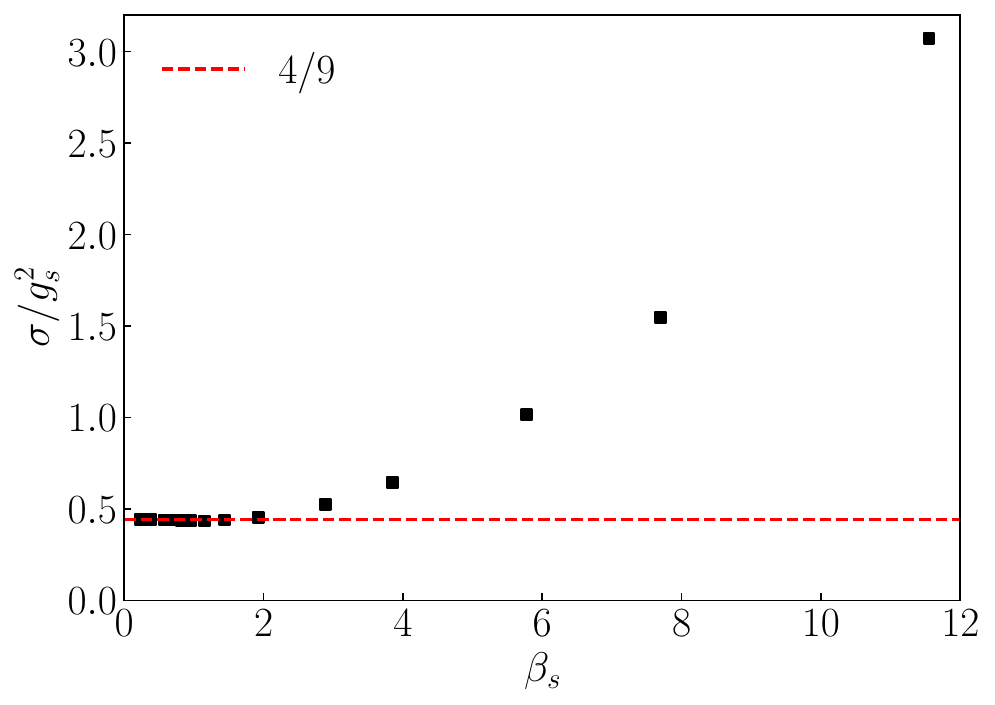}
\caption{String tension normalized by the effective coupling squared on the hexagonal lattice as a function of $\beta_s=N_c/g_s^2$. The dashed red line is the expectation in the strong coupling limit, which is set by the electric energy of a triplet link on the hexagonal lattice.}
\label{fig:sigma-g2-hex}
\end{figure}

\begin{figure}[h]
\centering
\includegraphics[width=0.9\linewidth]{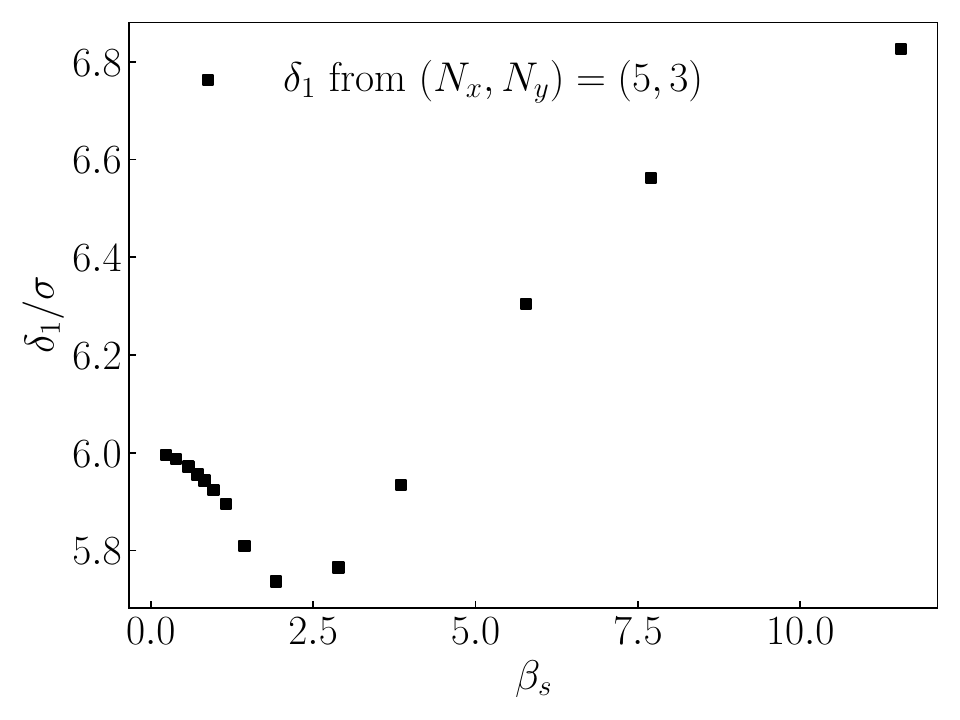}
\caption{Ground state energy gap $\delta_1$ normalized by the string tension as a function of $\beta_s$. The gap is calculated from the $5\times3$ hexagonal lattice.}
\label{fig:gap-sigma-g2-hex}
\end{figure}

The string tension extracted from a linear fit of $V_{Q\bar{Q}}$ calculated from different lattices ($N_y=3$, $N_x\in[3,4,5,6]$ for $g^2<1$, and $N_x\in[3,4,5]$ for $g^2 \geq 1$) with varying electric flux lengths is shown in Fig.~\ref{fig:sigma-g2-hex}, normalized by the effective coupling squared as a function of $\beta_s$. The dashed red line shows the value in the strong coupling limit, which is given by the electric energy per link, i.e., $4/3/n_l=4/9$ on the hexagonal lattice, where $n_l$ is defined in \eqref{eq:HKS}. The calculation results at strong coupling agree well with the expectation. The ground state energy gap $\delta_1$ calculated from the $5\times3$ hexagonal lattice and normalized by the string tension is shown in Fig.~\ref{fig:gap-sigma-g2-hex} as a function of $\beta_s$.  The normalized ground state energy gap exhibits a minimum at an intermediate coupling $g_s^2 \approx 1.5$.

\begin{figure}[t]
\centering
\includegraphics[width=0.9\linewidth]{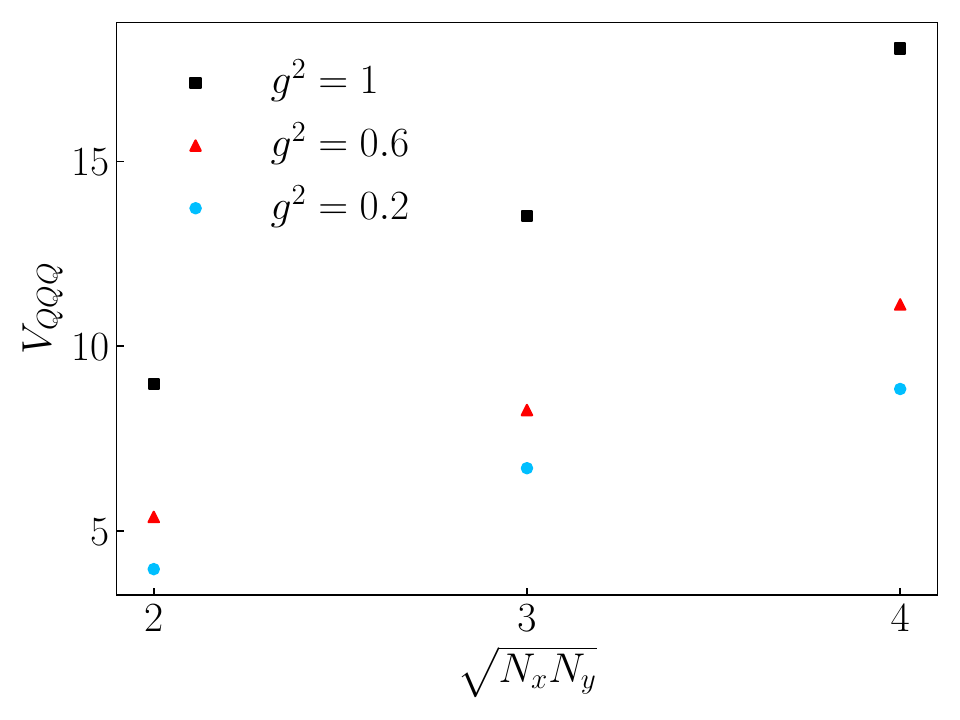}
\caption{$QQQ$ potential on hexagonal lattices of the approximate rectangle shape and sizes $N_x=N_y=2,3,4$ and for three couplings. The dependence on the system size is approximately linear.}
\label{fig:VQQQ}
\end{figure}

Finally, we study the potential energy associated with three static quarks. This can be investigated by injecting three fundamental electric fluxes on the boundary, as shown in Fig.~\ref{fig:QQQ}. This is special to the SU(3) gauge theory as three fundamental charges can form a singlet, which is absent in SU(2). The difference in the ground state energy of the flux-free and flux-injected lattices gives the $QQQ$ potential, i.e., $V_{QQQ}$. Figure~\ref{fig:VQQQ} shows its dependence on the lattice size specified by $\sqrt{N_xN_y}$ for three bare couplings. The results suggest a linear dependence.

\begin{figure}[h]
\centering
\includegraphics[width=0.9\linewidth]{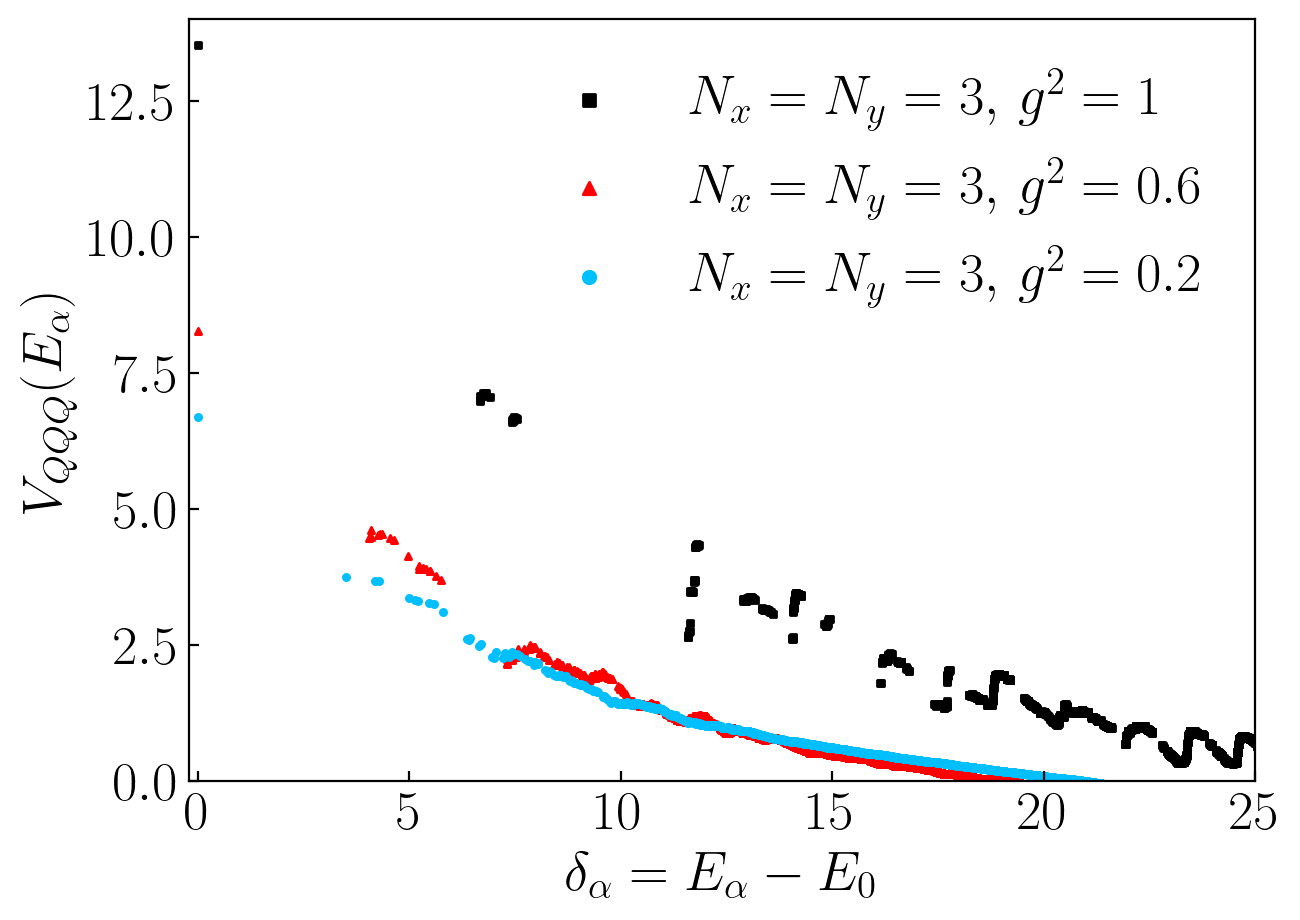}
\caption{Dependence of $V_{QQQ}$ on the excitation energy $E_\alpha$ of the flux-free $3\times3$ lattice with respect to its ground state energy. At high excitation energy, the $QQQ$ potential is significantly screened. The lattice system at the coupling $g^2=1$ is not quantum ergodic, which is why the black points do not follow a smooth curve.}
\label{fig:VQQQ-E}
\end{figure}

\begin{figure}[h]
\centering
\includegraphics[width=0.9\linewidth]{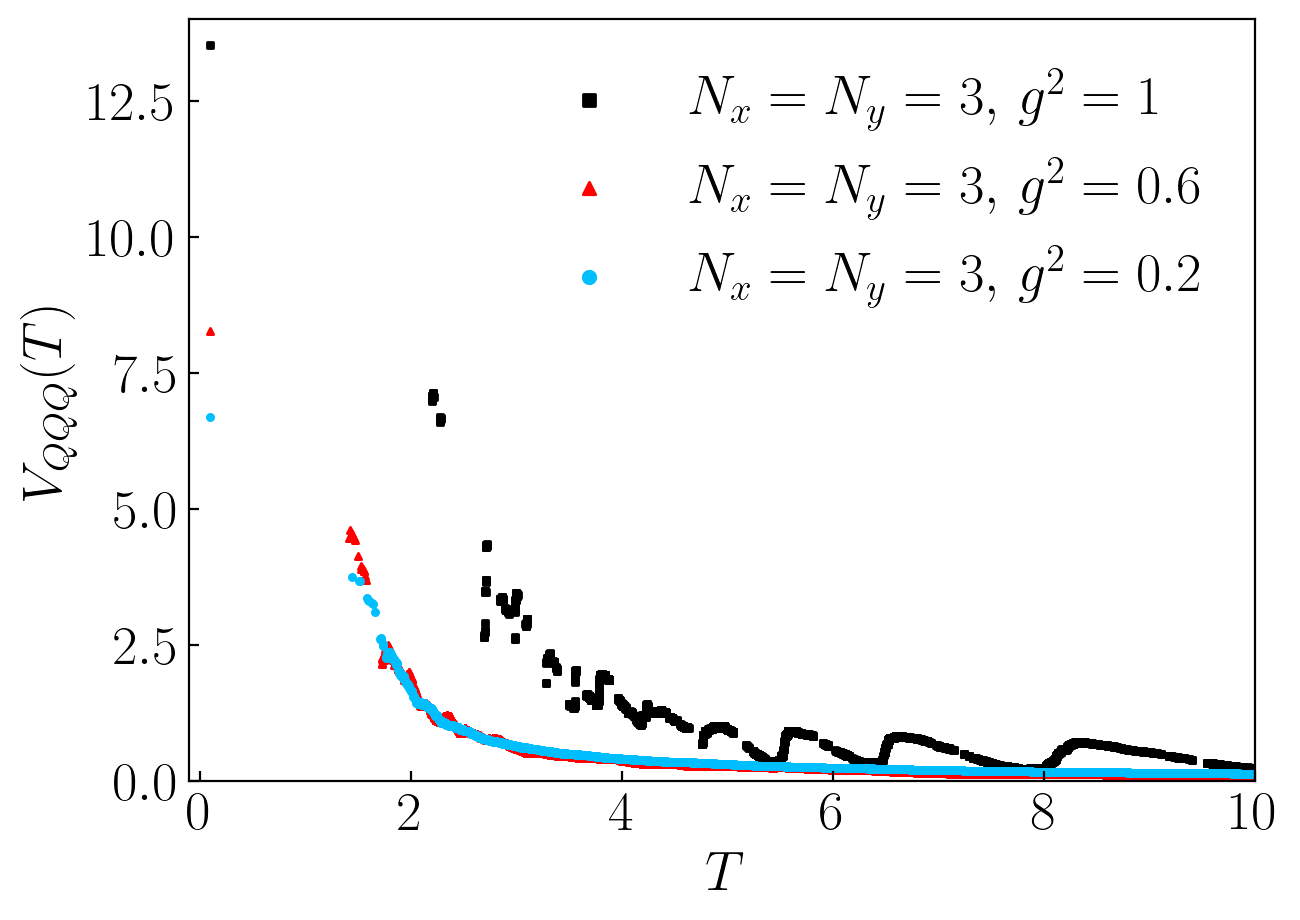}
\caption{Dependence of $V_{QQQ}$ on the canonical temperature of the flux-free $3\times3$ lattice. At high temperature, the $QQQ$ potential is significantly screened. The lattice system at the coupling $g^2=1$ is not quantum ergodic, which is why the black points do not follow a smooth curve.}
\label{fig:VQQQ-T}
\end{figure}

The dependence of $V_{QQQ}$ on the  excitation energy of the flux-free system and the corresponding canonical temperature is shown in Figs.~\ref{fig:VQQQ-E} and~\ref{fig:VQQQ-T}, respectively, for three couplings on the $3\times3$ lattice. The dependence is similar to the case of $V_{Q\bar{Q}}$. At high excitation energy and high temperature, the three-body string ``melts''. On such a small lattice, we already showed in Fig.~\ref{fig:gap-ratio-g2-hex} that the system at coupling $g^2=1$ is not quantum chaotic, and thus highly excited states do not behave thermally. This is why the black points in these two figures fluctuate strongly and do not follow a smooth curve as the other two sets of points. 

\section{Summary and Conclusions}
\label{sec:conclusions}

In this paper, we studied $(2+1)$D SU(3) gauge theory on trivalent lattices under the lowest nontrivial truncation in the electric basis, which includes the singlet, fundamental and antifundamental representations. Physical states can be specified by qutrits on each plaquette. Using the SU(3) 6j-symbols, we wrote down matrix elements for both the electric and magnetic (plaquette) terms in the lattice Hamiltonian. We briefly discussed the large $N_c$ limit of these matrix elements, as well as the insertions of static quarks on the lattice. 

By exactly diagonalizing the trivalent lattice Hamiltonians up to the size of 11 plaquettes, we studied level statistics of the system and identified the coupling regimes of quantum chaotic behavior at given lattice sizes. We also calculated the ground state energy gap up to the size of 18 plaquettes in total, which agrees with a previous Euclidean calculation result for the full SU(3) lattice gauge theory in the strong coupling regime. 

We also obtained the SU(3) string tension by injecting and extracting fundamental electric fluxes on the lattice boundary and calculating the ground state energy difference with and without the flux. The difference gives the potential energy between a static quark-antiquark pair and is shown to grow linearly with the distance between the source and sink. The potential energy decreases with the excitation energy of the system and the corresponding temperature, confirming the string ``melting'' phenomenon at finite temperature. Finally, we calculated the potential energy associated with three static quarks by injecting three fundamental fluxes into the system, a setup that is new compared to SU(2), and demonstrated its melting at finite temperature as well.

In future work, we plan to calculate matrix elements of physical operators and analyze various aspects of the eigenstate thermalization hypothesis, entanglement entropy, mutual information, antiflatness as the witness of the quantum magic resource, and dynamical quantities such as transport coefficients. We will also study the effects of higher SU(3) truncations and explore quantum simulation of these lattice systems. 

\begin{acknowledgements}
    B.M. acknowledges support by the U.S. Department of Energy, Office of Science (Grant DE-FG02-05ER41367) and by the National Science Foundation (Project PHY-2434506). X.Y. is supported by the U.S. Department of Energy, Office of Science, Office of Nuclear Physics, InQubator for Quantum Simulation (IQuS)\footnote{\url{https://iqus.uw.edu/}} under DOE (NP) Award Number DE-SC0020970 via the program on Quantum Horizons: QIS Research and Innovation for Nuclear Science.\footnote{\url{https://science.osti.gov/np/Research/Quantum-Information-Science}} This research used resources of the National Energy Research Scientific Computing Center (NERSC), a Department of Energy Office of Science User Facility using NERSC award NP-ERCAP0032083. This work was enabled, in part, by the use of advanced computational, storage and networking infrastructure provided by the Hyak supercomputer system at the University of Washington.
    We thank A. Ciavarella, D. Horn, and M. Savage for comments on a draft of the manuscript.
\end{acknowledgements}

\appendix

\section{$6j$-symbols for SU(3) and SU(4)}
\label{app:6j}

Numerical implementation of the lattice Hamiltonian in the minimal truncation requires knowledge of the eigenvalues of the quadratic Casimir operator and the relevant $6j$-symbols. Using the qudit notation: $\{{\bf 1},{\bf 3},{\bf \overline{3}}\} \to \{0,1,2\}$ for SU(3) and $\{{\bf 1},{\bf 4},{\bf 6},{\bf \overline{4}}\} \to \{0,1,2,3\}$ for SU(4), respectively, the Casimir eigenvalues are $C_2=\{0,\frac{4}{3},\frac{4}{3}\}$ for SU(3) and $C_2=\{0,\frac{15}{8},\frac{5}{2},\frac{15}{8}\}$ for SU(4). 

In the following two tables we present a list of the inverse squares of all relevant $6j$-symbols parametrized in terms of the first two entries ($a,b$) in the top row and the representation $r$ for the plaquette operator [$r=1$ for $U_P$ and $r=-1 \pmod{N_c}$ for $U_P^\dagger$]:
\[
\left\{\begin{array}{ccc}
a & b & -(a+b) \\
r & a+b+r & b+r \\
\end{array}\right\} \, (\rmod N_c).
\]
Table \ref{tab:6jsu3} lists the inverse squares of the $6j$-symbols for SU(3); Table \ref{tab:6jsu4} lists those for SU(4).

\begin{centering}
\begin{table}[h!]
\begin{tabular}{c|ccc}
& 0 & 1 & 2 \\
\hline
 0 & 3 & 9 & 3 \\
 1 & 9 & 9 & 9 \\
 2 & 9 & 9 & 9 \\
\end{tabular}
\qquad\qquad
\begin{tabular}{c|ccc}
& 0 & 1 & 2 \\
\hline
 0 & 3 & 3 & 9 \\
 1 & 9 & 9 & 9 \\
 2 & 9 & 9 & 9 \\
\end{tabular}
\caption{Inverse squares of the SU(3) $6j$-symbols for $r=1$ (left) and $r=-1\equiv_3 2$ (right). The row and column labels denote the values of $a$ and $b$. }
\label{tab:6jsu3}
\end{table}
\end{centering}

\begin{centering}
\begin{table}[h!]
\begin{tabular}{c|cccc}
& 0 & 1 & 2 & 3 \\
\hline
 0 & 4 & 24 & 24 & 4 \\
 1 & 16 & 36 & 16 & 16 \\
 2 & 24 & 24 & 24 & 24 \\
 3 & 16 & 16 & 36 & 16 \\
\end{tabular}
\qquad\qquad
\begin{tabular}{c|cccc}
& 0 & 1 & 2 & 3 \\
\hline
 0 & 4 & 4 & 24 & 24 \\
 1 & 16 & 16 & 36 & 16 \\
 2 & 24 & 24 & 24 & 24 \\
 3 & 16 & 16 & 16 & 36 \\
\end{tabular}
\caption{Inverse squares of the SU(4) $6j$-symbols for $r=1$ (left) and $r=-1\equiv_4 3$ (right). The row and column labels denote the values of $a$ and $b$.}
\label{tab:6jsu4}
\end{table}
\end{centering}

\section{Matrix elements of the magnetic energy operator for the plaquette chain}
\label{app:MPP'-chain}

For reference we list the matrix elements of the magnetic energy operator for the linear plaquette chain. As discussed, these depend on the states of the plaquettes adjacent to the four links defining a plaquette ($P_B,P_R,P_T,P_L$). In the absence of global electric flux through the plaquette chain, $P_B=P_T=|0\rangle$. Table \ref{tab:Hmag-chain} lists the matrix elements for this net flux-free case as $3\times 3$ matrices labeled by the quantum states of the plaquettes $P_L$ and $P_R$. These matrix elements agree with those listed in Appendix A of \cite{Ciavarella:2021nmj}.

\begin{centering}
\begin{table}[h!]
\begin{tabular}{c|ccc}
& 0 & 1 & 2 \\
\hline
 0\, & $\left(
 \begin{array}{ccc}
 0 & 1 & 1 \\
 1 & 0 & 1 \\
 1 & 1 & 0 \\
\end{array}
\right)$ & $\left(
\begin{array}{ccc}
 0 & \frac{1}{3} & \frac{1}{\sqrt{3}} \\
 \frac{1}{3} & 0 & \frac{1}{\sqrt{3}} \\
 \frac{1}{\sqrt{3}} & \frac{1}{\sqrt{3}} & 0 \\
\end{array}
\right)$ & $\left(
\begin{array}{ccc}
 0 & \frac{1}{\sqrt{3}} & \frac{1}{3} \\
 \frac{1}{\sqrt{3}} & 0 & \frac{1}{\sqrt{3}} \\
 \frac{1}{3} & \frac{1}{\sqrt{3}} & 0 \\
\end{array}
\right)$ \\
 & & & \\
 1\, & $\left(
\begin{array}{ccc}
 0 & \frac{1}{3} & \frac{1}{\sqrt{3}} \\
 \frac{1}{3} & 0 & \frac{1}{\sqrt{3}} \\
 \frac{1}{\sqrt{3}} & \frac{1}{\sqrt{3}} & 0 \\
\end{array}
\right)$ & $\left(
\begin{array}{ccc}
 0 & \frac{1}{9} & \frac{1}{3} \\
 \frac{1}{9} & 0 & \frac{1}{3} \\
 \frac{1}{3} & \frac{1}{3} & 0 \\
\end{array}
\right)$ & $\left(
\begin{array}{ccc}
 0 & \frac{1}{3 \sqrt{3}} & \frac{1}{3 \sqrt{3}} \\
 \frac{1}{3 \sqrt{3}} & 0 & \frac{1}{3} \\
 \frac{1}{3 \sqrt{3}} & \frac{1}{3} & 0 \\
\end{array}
\right)$ \\
 & & & \\
 2\, & $\left(
\begin{array}{ccc}
 0 & \frac{1}{\sqrt{3}} & \frac{1}{3} \\
 \frac{1}{\sqrt{3}} & 0 & \frac{1}{\sqrt{3}} \\
 \frac{1}{3} & \frac{1}{\sqrt{3}} & 0 \\
\end{array}
\right)$ & $\left(
\begin{array}{ccc}
 0 & \frac{1}{3 \sqrt{3}} & \frac{1}{3 \sqrt{3}} \\
 \frac{1}{3 \sqrt{3}} & 0 & \frac{1}{3} \\
 \frac{1}{3 \sqrt{3}} & \frac{1}{3} & 0 \\
\end{array}
\right)$ & $\left(
\begin{array}{ccc}
 0 & \frac{1}{3} & \frac{1}{9} \\
 \frac{1}{3} & 0 & \frac{1}{3} \\
 \frac{1}{9} & \frac{1}{3} & 0 \\
\end{array}
\right)$ \\
\end{tabular}
\caption{Matrix elements of the magnetic energy operator for the linear plaquette chain. The trivial diagonal elements are suppressed and the prefactor $1/g^2$ is omitted. The overall row and column labels indicate the quantum states $P_R,P_L$ of the adjacent plaquettes. Note that each $3\times 3$ matrix is symmetric, and the Table is symmetric with respect to the exchange $P_R \leftrightarrow P_L$.}
\label{tab:Hmag-chain}
\end{table}
\end{centering}

The matrix elements can be expressed in terms of the SU(3) T-, U-, and V-spin matrices as in Table~\ref{tab:Hmag-TUV}. 
\begin{centering}
\begin{table}[h!]
\begin{tabular}{c|ccc}
& 0 & 1 & 2 \\
\hline
0 & $T_x+U_x+V_x$ & 
$\frac{1}{3}T_x+\frac{U_x+V_x}{\sqrt{3}}$ &
$\frac{1}{3}U_x+\frac{T_x+V_x}{\sqrt{3}}$ \\
1 & $\frac{1}{3}T_x+\frac{U_x+V_x}{\sqrt{3}}$ &
$\frac{1}{9}T_x+\frac{1}{3}(U_x+V_x)$ &
$\frac{1}{3}V_x+\frac{T_x+U_x}{3\sqrt{3}}$ \\
2 & $\frac{1}{3}U_x+\frac{T_x+V_x}{\sqrt{3}}$ &
$\frac{1}{3}T_x+\frac{U_x+V_x}{\sqrt{3}}$ &
$\frac{1}{9}U_x+\frac{1}{3}(T_x+V_x)$ \\
\end{tabular}
\caption{Matrix elements of the magnetic energy operator for the linear plaquette chain expressed in terms of T-, U-, and V-spin matrices.}
\label{tab:Hmag-TUV}
\end{table}
\end{centering}

These SU(3) spin matrices can also be expressed in terms of the spin-1 matrices 
\begin{align}
&S_x = \frac{1}{\sqrt{2}}
\begin{pmatrix}
    0 & 1 & 0\\
    1 & 0 & 1\\
    0 & 1 & 0
\end{pmatrix} \,,\quad 
S_y = \frac{1}{\sqrt{2}}
\begin{pmatrix}
    0 & -i & 0\\
    i & 0 & -i\\
    0 & i & 0
\end{pmatrix}\,, \nn\\
&S_z = 
\begin{pmatrix}
    1 & 0 & 0\\
    0 & 0 & 0\\
    0 & 0 & -1
\end{pmatrix} \,.
\end{align}
With $S_\pm = S_x \pm i S_y$, we have
\begin{align}
    T_x &= \frac{(S_z^2+S_z)S_+}{2\sqrt{2}} + \frac{(1-S_z^2)S_-}{\sqrt{2}} \,, \nn\\
    V_x &= \frac{(S_z^2-S_z)S_-}{2\sqrt{2}} + \frac{(1-S_z^2)S_+}{\sqrt{2}} \,, \nn\\
    U_x &= \frac{S_+^2 + S_-^2}{2} \,.
\end{align}
Therefore, the SU(3) qutrit plaquette chain can be written as a spin-1 model. 

As discussed in Sec.~\ref{sec:string-chain} a global fundamental electric flux through the plaquette chain by choosing $P_B$ or $P_T$ in the ${\bf 3}$ or $\bf\overline{3}$ representation. Here we present the plaquette matrix element for $P_B=|0\rangle$, $P_T=|2\rangle$ (using the qutrit notation) in Table \ref{tab:Hmag-chain-flux}. This corresponds to fundamental electric flux being injected at the top right corner and extracted at the top left corner of the chain.

\begin{centering}
\begin{table}[h!]
\begin{tabular}{c|ccc}
& 0 & 1 & 2 \\
\hline
 0\, & $\left(
 \begin{array}{ccc}
 0 & \frac{1}{\sqrt{3}} & \frac{1}{3} \\
 \frac{1}{\sqrt{3}} & 0 & \frac{1}{\sqrt{3}} \\
 \frac{1}{3} & \frac{1}{\sqrt{3}} & 0 \\
\end{array}
\right)$ & $\left(
\begin{array}{ccc}
 0 & \frac{1}{3} & \frac{1}{3} \\
 \frac{1}{3} & 0 & \frac{1}{3} \\
 \frac{1}{3} & \frac{1}{3} & 0 \\
\end{array}
\right)$ & $\left(
\begin{array}{ccc}
 0 & \frac{1}{\sqrt{3}} & \frac{1}{3} \\
 \frac{1}{\sqrt{3}} & 0 & \frac{1}{\sqrt{3}} \\
 \frac{1}{3} & \frac{1}{\sqrt{3}} & 0 \\
\end{array}
\right)$ \\
 & & & \\
 1\, & $\left(
\begin{array}{ccc}
 0 & \frac{1}{3} & \frac{1}{3} \\
 \frac{1}{3} & 0 & \frac{1}{3} \\
 \frac{1}{3} & \frac{1}{3} & 0 \\
\end{array}
\right)$ & $\left(
\begin{array}{ccc}
 0 & \frac{1}{3 \sqrt{3}} & \frac{1}{3} \\
 \frac{1}{3 \sqrt{3}} & 0 & \frac{1}{3 \sqrt{3}} \\
 \frac{1}{3} & \frac{1}{3 \sqrt{3}} & 0 \\
\end{array}
\right)$ & $\left(
\begin{array}{ccc}
 0 & \frac{1}{3} & \frac{1}{3} \\
 \frac{1}{3} & 0 & \frac{1}{3} \\
 \frac{1}{3} & \frac{1}{3} & 0 \\
\end{array}
\right)$ \\
 & & & \\
 2\, & $\left(
\begin{array}{ccc}
 0 & \frac{1}{\sqrt{3}} & \frac{1}{3} \\
 \frac{1}{\sqrt{3}} & 0 & \frac{1}{\sqrt{3}} \\
 \frac{1}{3} & \frac{1}{\sqrt{3}} & 0 \\
\end{array}
\right)$ & $\left(
\begin{array}{ccc}
 0 & \frac{1}{3} & \frac{1}{3} \\
 \frac{1}{3} & 0 & \frac{1}{3} \\
 \frac{1}{3} & \frac{1}{3} & 0 \\
\end{array}
\right)$ & $\left(
\begin{array}{ccc}
 0 & \frac{1}{\sqrt{3}} & \frac{1}{3} \\
 \frac{1}{\sqrt{3}} & 0 & \frac{1}{\sqrt{3}} \\
 \frac{1}{3} & \frac{1}{\sqrt{3}} & 0 \\
\end{array}
\right)$ \\
\end{tabular}
\caption{Matrix elements of the magnetic energy operator for the linear plaquette chain in the presence of global fundamental electric flux runing from right to left along the top of the lattice. The trivial diagonal elements are suppressed and the prefactor $1/g^2$ is omitted. The overall row and column labels indicate the quantum states $P_R,P_L$ of the adjacent plaquettes. Note that each $3\times 3$ matrix is symmetric, and the Table is symmetric with respect to the exchange $P_R \leftrightarrow P_L$.}
\label{tab:Hmag-chain-flux}
\end{table}
\end{centering}

\FloatBarrier

\bibliographystyle{apsrev4-1}
\bibliography{SU3_refs}

\end{document}